\documentclass[useAMS, usenatbib, usegraphcx]{mnras}
\usepackage[english]{babel}

\usepackage{acronym}
\usepackage{amsmath}
\usepackage{amssymb}
\usepackage{bm}
\usepackage{color}
\usepackage{graphicx}
\usepackage{latexsym}
\usepackage{multicol, multirow}
\usepackage{stmaryrd}
\usepackage[normalem]{ulem}
\usepackage{url}

\usepackage{tensind}
\tensordelimiter{?}
\tensorformat{lrc}

\newcommand{\dd}{\mathrm{d}}
\newcommand{\eff}{\mathrm{eff}}

\newcommand{\cf}{\textit{cf.}}
\newcommand{\ie}{\textit{i.e.}}
\newcommand{\eg}{\textit{e.g.}}

\begin{document}

\newcommand{\cqg}{Class.~Quantum~Grav.}
\newcommand{\jcomp}{J.~Comput.~Phys.~}
\newcommand{\pr}{Phys.~Rev.}

\title{Dynamical Mass Ejection from Binary Neutron Star Mergers}
\author[D.~Radice et al.]{%
David Radice$^{1}$,
Filippo Galeazzi$^{2}$,
Jonas Lippuner$^{1}$,
Luke F.~Roberts$^{1,\ast}$,
\newauthor
Christian D.~Ott$^{1}$,
and Luciano Rezzolla$^{2,3}$ \\
$^1$TAPIR, Walter Burke Institute for Theoretical Physics, MC 350-17,
California Institute of Technology, Pasadena, CA 91125, USA \\
$^2$Institute for Theoretical Physics, Max-von-Laue Str. 1, 60438,
Frankfurt, Germany \\
$^3$Frankfurt Institute for Advanced Studies, Ruth-Moufang-Str. 1,
60438 Frankfurt, Germany\\
$^\ast$NASA Einstein Fellow
}

\maketitle
\begin{abstract}
We present fully general-relativistic simulations of binary neutron star mergers
with a temperature and composition dependent nuclear equation of state. We study
the dynamical mass ejection from both quasi-circular and dynamical-capture
eccentric mergers. We systematically vary the level of our treatment of the
microphysics to isolate the effects of neutrino cooling and heating and we
compute the nucleosynthetic yields of the ejecta. We find that eccentric
binaries can eject significantly more material than quasi-circular binaries and
generate bright infrared and radio emission. In all our simulations the outflow
is composed of a combination of tidally- and shock-driven ejecta, mostly
distributed over a broad $\sim 60^\circ$ angle from the orbital plane, and, to a
lesser extent, by thermally driven winds at high latitudes.  Ejecta from
eccentric mergers are typically more neutron rich than those of quasi-circular
mergers.  We find neutrino cooling and heating to affect, quantitatively and
qualitatively, composition, morphology, and total mass of the outflows. This is
also reflected in the infrared and radio signatures of the binary. The final
nucleosynthetic yields of the ejecta are robust and insensitive to input physics
or merger type in the regions of the second and third r-process peaks. The
yields for elements on the first peak vary between our simulations, but none of
our models is able to explain the Solar abundances of first-peak elements
without invoking additional first-peak contributions from either neutrino and
viscously-driven winds operating on longer timescales after the mergers, or from
core-collapse supernovae.
\end{abstract}

\section{Introduction}
Binary neutron star (BNS)\acused{BNS}\acused{NS} and black-hole neutron-star
(BHNS) \acused{BH}\acused{BHNS} mergers can result in the dynamical ejection of
\ac{NS} matter due to tidal torques and/or, in the \ac{BNS} case, shocks during
the merger. This neutron-rich material has long been proposed as a possible
origin for the elements with atomic mass number $A\gtrsim 120$
\citep{lattimer:74, meyer:89, eichler:89, freiburghaus:99}. As the ejecta expand
and cool they realize the right conditions for the activation of the so-called
rapid neutron capture process (\emph{r-process}), synthesizing neutron-rich
nuclei.

This scenario has received renewed attention in recent years. On the one hand,
detailed calculations have shown that core-collapse supernovae, once considered
to be the main candidate for the production of the r-process elements, do not
appear to host the right conditions to create the most neutron-rich nuclei
\citep[\eg,][]{roberts:10, hudepohl:10, fischer:10}. On the other hand, the most
recent \ac{GR} simulations of the mass ejection occurring during \ac{BNS}
\citep{hotokezaka:13, bauswein:13, wanajo:14, sekiguchi:15, kastaun:15,
palenzuela:15, foucart:15b} and \ac{BHNS} \citep[\eg,][]{foucart:14, kyutoku:15}
mergers, as well as in the later evolution of the post-merger remnant
\citep[\eg,][]{dessart:09, metzger:14, fernandez:15, foucart:15a, just:15a,
martin:15}, suggest that mergers are likely to eject a large amount of r-process
material \citep{rosswog:99, korobkin:12, wanajo:14, just:15a, martin:15}.

In the past, simple models of galactic chemical evolution found issues with the
evolution of the r-process abundances under the assumption that compact object
binary mergers are the primary source of the r-process. The large amount of
material ejected per event predicts a larger spread in r-process enrichment of
metal poor halo stars than what is observed, when using simple models of
galactic chemical evolution \citep{qian:00, argast:04}.  Additionally, the delay
time between binary formation and merger can result in r-process nucleosynthesis
only occurring at somewhat higher metallicity than it is observed to begin at
\citep{argast:04}.  In more recent models of galactic chemical evolution
performed within cosmological zoom-in simulations \citep{shen:15,
van-de-voort:15}  or by accounting for accretion of sub-halos into the Milky Way
halo \citep{ishimaru:15} these problems seem to be mitigated and better
agreement with the observed distribution of r-process elements in the Milky Way
is found. However, the results of these models are sensitive to numerical
resolution and to their treatment of mixing.

On the other hand, the idea that compact binary mergers could be the site of the
r-process is also tentatively supported by the recent discovery of an infrared
transient associated with Swift \ac{SGRB} 130603B \mbox{\citep{berger:13,
tanvir:13}} that might be explained by the radioactive decay of by-products of
the r-process in a \emph{macronova} (sometimes also called \emph{kilonova})
\citep[\eg,][]{Li:98, kulkarni:05, metzger:10, roberts:11, kasen:13, barnes:13,
tanaka:13, rosswog:14, grossman:14, lippuner:15}. Other evidence includes the
Solar system abundance of $^{244}$Pu \citep{wallner:15, hotokezaka:15} and
recent observations of r-process enriched stars in a metal-poor ultra-faint
dwarf galaxy \citep{ji:15}. Both of these observations suggest that r-process
elements might be preferentially produced in rare/high-yield events such as
mergers instead of common/low-yield occurrences such as core-collapse
supernovae.

Beside powering macronovae, the outflows from \ac{BNS} and \ac{BHNS} mergers
could also generate radio flares over timescales of months to years as their
kinetic energy is deposited in the interstellar medium \citep{nakar:11}, and
could explain the extended X-ray emission observed in some \acp{SGRB}
\citep{rezzolla:14, ciolfi:14}. Mergers are also loud \ac{GW} sources
and one of the targets for the nascent field of \ac{GW} astronomy
\citep{sathyaprakash:09}, recently inaugurated with the detection of \acp{GW}
from a pair of merging \acp{BH} \citep{gwdetection}. \ac{BNS} mergers are
promising sources for ground-based laser-interferometer detectors such as
Advanced LIGO \citep{advligo}, Advanced Virgo \citep{advvirgo}, and KAGRA
\citep{kagra}.  Finally, \ac{BNS} and \ac{BHNS} mergers are also thought to
create the central engines of \ac{SGRB} \citep{nakar:07, berger:14, rosswog:15}.
This makes \ac{BNS} and \ac{BHNS} mergers ideal candidates for multi-messenger
astronomy \citep{metzger:12, nissanke:13} and motivates the systematic study of
their observational signatures.

Most previous studies of \ac{BNS} and \ac{BHNS} mergers focused on the case of
``primordial'' binaries (\ie, formed in an already bound state) which merge
under the effect of \ac{GW} losses at zero eccentricity. However,
high-eccentricity mergers might also occur in dense stellar environments such as
\acp{GC}. About $10\%$ of all observed \acp{SGRB} show offsets larger than $20$
kiloparsecs from the bulge of their host galaxies \citep{berger:05, berger:14}.
In comparison, no long-GRBs have been found with an offset of more than 10
kiloparsecs from their host galaxies \citep{berger:14}.  These offsets could be
the result of kicks imparted to the binaries at birth, or they could be
explained by $10\%$ of SGRB progenitors being located in \acp{GC} around their
host galaxies \citep{berger:14}. \acp{GC} are dense environments (especially
when they undergo the core collapse phase) in which tidal captures, collisions,
and dynamically formed binaries from two- or three-body interactions become a
more viable channel for \ac{BNS} and \ac{BHNS} mergers than in the galactic
stellar field \citep{lee:10}.

Current estimates for the rates of eccentric \ac{BNS} mergers in \acp{GC} are
unfortunately plagued by many uncertainties stemming from unknown properties of
\acp{GC}, such as the number of \acp{NS} in the \ac{GC} core \citep{murphy:11}.
Recently, \citet{tsang:13} reviewed the estimates for binary collisions and
tidal captures in \ac{GC} and arrived at a conservative estimate of $0.5 \;
\text{yr} ^{-1} \; \text{Gpc}^{-3}$. However, \citet{tsang:13} leaves open the
possibility that neglected contributions, such as the interaction of a single
\ac{NS} with a binary system or rarer binary-binary interactions, might increase
these rates.  In any case, even under the most optimistic assumptions,
dynamical-capture \ac{BNS} mergers would account for only a few percent of the
total \ac{BNS} merger rate.

Nevertheless, eccentric mergers might eject significantly larger amounts of mass
than non-eccentric, quasi-circular, mergers \citep{east:12b, rosswog:13}. As a
consequence, they could be contributing an important fraction of the overall
r-process element abundances, despite being rarer. For the same reason,
eccentric \ac{BNS} mergers would also give birth to particularly bright \ac{EM}
emissions in the infrared and radio bands compared with quasi-circular binaries.
Additionally, eccentric \ac{BNS} mergers produced by dynamical captures in dense
stellar environments can potentially produce r-process nuclei at very low
metallicity and account for the r-process enhancements seen in some carbon
enhanced metal poor stars \citep{ramirez-ruiz:15}.

To date, only few \ac{GR} studies considered eccentric \ac{BNS} \citep{gold:12,
east:12b, paschalidis:15, east:15b} and \ac{BHNS} \citep{stephens:11, east:12a,
east:15a} mergers. Eccentric mergers were also considered in Newtonian physics
by \citet{lee:10} and  \citet{rosswog:13}. All of these studies, apart from the
Newtonian simulations of \citet{rosswog:13}, employed idealized equations of
state. No previous study considered eccentric mergers in full-\ac{GR}, with a
microphysical temperature- and composition-dependent equation of state, and with
the inclusion of neutrino emission and absorptions. GR, microphysical \ac{EOS},
and neutrinos are three ingredients that are necessary to accurately model the
mass ejection and the nucleosynthetic yields from these events. Neutrino
transport, in particular, has been recently suggested to have an important role
in shaping the composition of the ejecta by \citet{wanajo:14, sekiguchi:15,
foucart:15a, foucart:15b}.  This, in turn, might affect the nucleosynthetic
yields \citep{wanajo:14, goriely:15} and, possibly, the properties of the
macronova emission \citep{metzger:14, lippuner:15}.

The goal of this paper is to develop a comprehensive understanding of the
physics driving dynamical mass ejection in \ac{BNS} mergers. On the one hand, we
quantify the impact of weak reactions and neutrino radiation on the properties
of the ejecta, including composition and geometry, extending the works of
\citet{wanajo:14, sekiguchi:15, foucart:15a, foucart:15b}. On the other hand, we
study, for the first time in full-\ac{GR} and including a microphysical equation
of state and neutrino cooling/heating, the mass ejection in eccentric mergers.

Toward these goals we perform a series of full-\ac{GR} simulations of merging
\ac{BNS} in eccentric and quasi-circular orbits. For the eccentric mergers, we
consider $5$ possible configurations leading to a variety of different outcomes,
including prompt and delayed \ac{BH} formation, while all our quasi-circular
simulations show delayed \ac{BH} formation. We systematically vary the level of
sophistication of our microphysical description to isolate the effects of local
weak reactions (mainly captures on neutrons) and neutrino irradiation from the
merger remnant.

The remainder of this paper is organized as follows. We discuss our numerical
models, simulation methods, and initial data in Section \ref{sec:method}. We
present our main results in Section~\ref{sec:results}, while Section
\ref{sec:conclusions} is dedicated to discussion and conclusions. Appendix
\ref{sec:M0} describes our treatment of neutrino radiation. Finally, Appendix
\ref{sec:coupling} contains some non-standard implementation details of the
operator split technique we use to treat the weak source/sink terms in the
hydrodynamics equations.

Unless otherwise specified, we use a system of units such that $c = G = M_\odot
= 1$, where $c$ is the speed of light in vacuum, $G$ is the gravitational
constant,  and $M_\odot$ is the mass of the Sun. We use Einstein's convention of
summation over repeated indices. Latin indices run over $1,2,3$, while Greek
indices run over $0,1,2,3$. The spacetime metric signature we adopt is
$(-,+,+,+)$.

\section{Numerical Model}
\label{sec:method}
\subsection{General-Relativistic Hydrodynamics}
\label{sec:grhydro}
We simulate the merging \acp{NS} using the equations of \ac{GR} hydrodynamics
\citep[\eg,][]{Rezzolla_book:2013}.  We evolve the equation of baryon number
conservation
\begin{equation}\label{eq:continuity}
  \nabla_\alpha (n\, u^\alpha) = 0\,,
\end{equation}
where $n$ is the baryon-number density and $u^\alpha$ is the fluid
four-velocity, together with the equations describing conservation of energy and
momentum,
\begin{equation}\label{eq:euler}
  \nabla_\beta T^{\alpha\beta} = \Psi^\alpha\,.
\end{equation}
\ac{NS} matter is treated as a perfect fluid and thus with energy-momentum
tensor
\begin{equation}\label{eq:tmunu}
  T^{\alpha\beta} = \rho\, h\, u^\alpha\, u^\beta + p\, g^{\alpha\beta}\,,
\end{equation}
where $\rho$ is the rest-mass density,
\begin{equation}\label{eq:rmd}
  \rho = m_b\, n\,,
\end{equation}
$m_b$ is the fiducial baryon mass, $h = 1 + \epsilon + p/\rho$ is the specific
enthalpy, $\epsilon$ is the specific internal energy, $p$ is the pressure, and
$g^{\alpha\beta}$ is the spacetime metric. Finally, $\Psi^\alpha$ is a source
term used to model weak interactions (more on this in Sect.  \ref{sec:leakage}).

Equations \eqref{eq:continuity} and \eqref{eq:euler} are closed by means of an
\ac{EOS} $p = p(n, \epsilon, Y_e)$, where $Y_e$ is the proton (or electron)
fraction. For this work we adopt the equation of state of \cite{Lattimer:91},
with nuclear compressibility parameter $K = 220\ \mathrm{MeV}$. This equation of
state is broadly consistent with observations, but falls outside the favored
region of microscopic neutron matter calculations \citep[\eg,][]{fischer:14}. It
predicts a maximum non-rotating \ac{NS} gravitational mass of $2.05\ M_\odot$
and a $1.4\ M_\odot-$\ac{NS} circumferential radius of $12.8\ \mathrm{km}$.

In some of the simulations we neglect weak interactions and we assume the proton
number $n_e = n Y_e$ to be conserved. In this case, we also set $\Psi^\alpha$ in
equation \eqref{eq:euler} to zero. In most simulations, however, we do not
assume $n_e$ to be conserved, but we evolve the composition taking into account
weak interactions,
\begin{equation}\label{eq:composition}
  \nabla_\alpha (n_e\, u^\alpha) = R\,,
\end{equation}
where $R$ is the net lepton number emission/absorption rate per unit volume in
the fluid rest-frame. In this case, the energy-momentum source terms read
\begin{equation}\label{eq:cooling}
  \Psi^\alpha = Q\, u^\alpha\,,
\end{equation}
where $Q$ is the net neutrino cooling/heating rate per unit volume in the fluid
rest-frame. The way we compute $R$ and $Q$ is described in Sect. \ref{sec:leakage}.

Equations \eqref{eq:tmunu} and \eqref{eq:composition} are discretized in
flux-conservative form using the \texttt{WhiskyTHC} code \citep{radice:12b,
radice:14a, radice:15b}. \texttt{WhiskyTHC} implements both finite-volume and
high-order finite-differencing high-resolution shock-capturing methods. For the
simulations presented in this work, we use the high-order MP5 primitive
reconstruction scheme \citep{suresh:97} in combination with 2nd order numerical
fluxes computed with the HLLE flux formula \citep{HLLE:88}. In all simulations
we employ the positivity preserving limiter presented in \citet{radice:14b},
which minimizes errors related to the numerical density floor.

\texttt{WhiskyTHC} directly evolves the proton $n_e$ and neutron $n-n_e$ number
densities thereby guaranteeing the local conservation of both species. In
addition, \texttt{WhiskyTHC} also implements a variant of the \ac{CMA} method
originally proposed by \citet{plewa:99}. This scheme minimizes advection errors
near strong density and compositional gradients, such as at the surface of a
\ac{NS}. We find the use of the \ac{CMA} scheme to be crucial for quasi-circular
binaries that otherwise accumulate significant advection errors close to the
surface during their inspiral. On the other hand, the \ac{CMA} method does not
seem to be critical for eccentric binaries. The latter have a relatively short
approaching phase and, in the case of models undergoing multiple encounters,
generate a dense, well-resolved, ``envelope'' after their first encounter (see
Sect. \ref{sec:results.dynamics}). For this reason, we employ the \ac{CMA}
scheme only for the quasi-circular binaries. However, we verify the robustness
of our results by simulating one of the eccentric models using the \ac{CMA}
scheme.

Finally, we evolve the spacetime with the \mbox{BSSNOK} formulation of the
Einstein equations \citep{Nakamura87, Shibata95, Baumgarte99}, using the
\texttt{Mclachlan} code \citep{Brown:2008sb}, which is part of the
\texttt{Einstein Toolkit} \citep{Loffler:2011ay}, and a fourth-order accurate
finite-difference scheme.

\subsection{Neutrino Treatment}
\label{sec:leakage}
Our neutrino treatment is based on a so-called gray (energy-averaged) leakage
scheme, which is a parametrized neutrino cooling scheme that has been widely
used for both core-collapse supernovae and \ac{BNS} simulations
\citep[\eg,][]{vanriper:81, ruffert:96, rosswog:03, oconnor:10, perego:15}. Our
leakage scheme is an evolution of the scheme presented in \citet{galeazzi:13}.
It follows very closely the method used in \citet{ruffert:96}, with the
additional simplification that we use the local thermodynamical equilibrium
chemical potential for the neutrinos while computing opacities as done in
\citet{rosswog:03}. Finally, for the calculation of the optical depth, we adopt
the prescription presented in \citet{neilsen:14}, which is well suited for
complex geometries as the ones encountered in \ac{BNS} mergers.

Here, we limit our discussions to the aspects of our leakage scheme that differs
from theirs. The basic idea of the leakage scheme is to compute a series of
effective emissivities $R^{\eff}$ and $Q^{\eff}$ for electron neutrinos $\nu_e$,
anti-electron neutrinos $\bar{\nu}_e$ and the heavy-lepton neutrinos, which we
collectively label as $\nu_x$. These rates are reduced with respect to the
intrinsic emissivities in a way that mimics the diffusion of radiation from the
high optical depth region. In addition to this, our scheme also estimates
heating and lepton-deposition from the absorption of free-streaming neutrinos.
The inclusion of neutrino absorption is motivated by the recent works of
\citet{wanajo:14, sekiguchi:15, foucart:15a, foucart:15b}, which showed, using
gray two-moment schemes \citep{shibata:12}, that neutrino absorption can alter
the composition of the ejecta. The source terms appearing in
\eqref{eq:composition} and \eqref{eq:cooling} are then computed as
\begin{equation}\label{eq:composition.rhs}
  R = (\kappa_{\nu_e} n_{\nu_e} - \kappa_{\bar{\nu}_e} n_{\bar{\nu}_e}) -
      (R^\eff_{\nu_e} + R^\eff_{\bar{\nu}_e} )
\end{equation}
and
\begin{equation}\label{eq:cooling.rhs}
  Q = (\kappa_{\nu_e} n_{\nu_e} E_{\nu_e}
    + \kappa_{\bar{\nu}_e} n_{\bar{\nu}_e} E_{\bar{\nu}_e})
    - (Q^\eff_{\nu_e} + Q^\eff_{\bar{\nu}_e} + Q^\eff_{\nu_x})\,,
\end{equation}
where the $\kappa$'s are the absorption opacities and $n_{\nu_e}$,
$n_{\bar{\nu}_e}$ are the free-streaming neutrino and anti-neutrino number
densities in the fluid rest-frame. Finally, $E_{\nu_e}$ and $E_{\bar{\nu}_e}$
are the average energies of the free-streaming neutrinos in the fluid
rest-frame.

To compute $n_{\nu_e}$, $n_{\bar{\nu}_e}$, $E_{\nu_e}$, and $E_{\bar{\nu}_e}$,
we evolve the zeroth moment (M0) of the free-streaming neutrino distribution
function on a set of individual radial rays. For that, we use a closure adapted
to the post-merger geometry discussed in detail in Appendix~\ref{sec:M0}.  Our
scheme is simpler than the two-moment gray method recently used by
\citet{wanajo:14, sekiguchi:15, foucart:15a, foucart:15b}. However, since it
tracks both neutrino density and average energies, it allows us to model a
number of important effects that cannot be easily incorporated into gray
schemes.  In particular, our scheme takes gravitational redshift, velocity
dependence and non local-thermodynamical equilibrium effects into account,
albeit with some major approximations. More details of our radiation transport
scheme are given in Appendix~\ref{sec:M0}, while some technical details of the
coupling with the hydrodynamics are discussed in Appendix~\ref{sec:coupling}.
Note that heating in the M0 code is switched-on only shortly before the merger,
because our prescription is not suitable for the phase when the two \ac{NS}s are
still separated and heating is obviously not relevant. After \ac{BH} formation,
we also excise the area inside and close to the apparent horizon in the M0
scheme. In most simulations, with the exception of the \texttt{M0\_QC}, we do
not excise the hydrodynamic variables, but follow the approaches described in
\citet{baiotti:06}, for the metric evolution, and in \citet{galeazzi:13}, for
the \ac{EOS} and primitive recovery routines.

\subsection{Initial Data and Grid Setup}

\begin{table*}
  \caption{Summary of key results for all models. We report the model name, the
  (Newtonian) periastron radius $r_p$, the total ejected mass $M_{\mathrm{ej}}$,
  the mass-averaged proton fraction $\langle Y_e \rangle$, specific entropy per
  baryon $\langle s \rangle$, and asymptotic velocity of the ejecta, $\langle
  v_\infty \rangle$, measured on a sphere with coordinate radius $r = 200\
  M_\odot \simeq 295\ \mathrm{km}$. We also report the total kinetic energy of
  the ejecta $E_{\mathrm{kin}}$, the macronova peak time $t_{\mathrm{peak}}$
  (equation \ref{eq:time_rad_peak}) luminosity $L$ (equation
  \ref{eq:lum_rad_peak}), and effective temperature $T$ (equation
  \ref{eq:temp_rad_peak}), as well as the deceleration time of the ejecta
  $t_{\mathrm{dec}}$ (equation \ref{eq:time_dec}) and the radio fluence $F_\nu$
  (\ie, the flux density per unit frequency) during $t_{\mathrm{dec}}$ (equation
  \ref{eq:obs_flux}). In the case of simulations where we observe
  \ac{BH} formation, we also include the remnant torus mass $M_{\mathrm{torus}}$
  at $1\ \mathrm{ms}$ after the formation of the apparent horizon. Simulations
  with names with prefix \texttt{LK} do not account for neutrino re-absorption,
  while runs with prefix \texttt{M0} include neutrino re-absorption using the
  scheme discussed in Appendix~\ref{sec:M0}.  Simulations with names prefixed by
  \texttt{HY} neglect all weak interactions.  Names with suffix \texttt{QC}
  represent models constructed with quasi-circular initial data, while names
  with suffix \texttt{RPX} represent parabolic models having (Newtonian)
  periastron radius of $X\ M_\odot$.}
  \label{tab:models}
  \begin{center}
  \resizebox{\textwidth}{!}{
  \begin{tabular}{lcccccccccccc}
  \hline\hline
    \multirow{2}{*}{Model} &
    $r_p$ &
    $M_{\mathrm{torus}}$ &
    $M_{\mathrm{ej}}$ &
    $\langle Y_e \rangle$ &
    $\langle s   \rangle$ &
    $\langle v_\infty \rangle$ &
    $E_{\mathrm{kin}}\ $ &
    $t_{\mathrm{peak}}$ &
    $L$ &
    $T$ &
    $t_{\mathrm{dec}}$ &
    $F_\nu$ \\
    &
    $[\mathrm{km}]$ &
    $[10^{-2}\ M_\odot]$ &
    $[10^{-2}\ M_\odot]$ &
    & $[k_b]$ &
    $[10^{-1}\ c]$ &
    $[10^{51}\ \mathrm{erg}]$ &
    $[\mathrm{days}]$ &
    $[10^{41}\ \mathrm{erg}\ \mathrm{s}^{-1}]$ &
    $[10^3 K]$ &
    $[\mathrm{years}]$ &
    $[\mathrm{mJy}]$ \\
  \hline
    \texttt{HY\_QC}$^a$ &

      $45.00$ & $13.48$ & $\phantom{0}0.49$ & $0.04$ & $18.8$ & $1.5$ & $0.125$
      & $\phantom{0}2.8$ & $0.25$ & $2.4$ & $\phantom{0}9.9$ & $0.027$ \\
    \texttt{LK\_QC}$^a$

      & $45.00$ & $\phantom{0}5.45$ & $\phantom{0}0.16$ & $0.15$ & $14.4$ & $1.9$ & $0.065$
      & $\phantom{0}1.4$ & $0.20$ & $2.8$ & $\phantom{0}5.3$ & $0.028$ \\
    \texttt{M0\_QC}$^a$

      & $45.00$ &  $\phantom{0}7.40$ &  $\phantom{0}0.17$ &  $0.17$ &  $14.8$ &  $1.6$ &  $0.053$
      &  $\phantom{0}1.6$ &  $0.18$ &  $2.8$ &  $\phantom{0}6.5$ &  $0.015$ \\
  \hline
    \texttt{LK\_RP0}$^b$

      & $\phantom{0}0.00$ & $\phantom{0}0.00$ & $-$ & $-$ & $-$ & $-$ & $-$
      & $-$ & $-$ & $-$ & $-$ & $-$ \\
  \hline
    \texttt{HY\_RP5}    &

      $\phantom{0}7.38$ & $\phantom{0}0.17$ & $\phantom{0}0.04$ & $0.05$ & $28.4$ & $4.1$ & $0.078$
      & $\phantom{0}0.5$ & $0.20$ & $3.3$ & $\phantom{0}1.5$ & $0.285$ \\
    \texttt{LK\_RP5}

      & $\phantom{0}7.38$ & $\phantom{0}0.17$ & $\phantom{0}0.02$ & $0.20$ & $17.1$ & $3.3$ & $0.025$
      & $\phantom{0}0.4$ & $0.14$ & $3.8$ & $\phantom{0}1.5$ & $0.051$ \\
    \texttt{M0\_RP5}
      & $\phantom{0}7.38$ &  $\phantom{0}0.16$ &  $\phantom{0}0.02$ &  $0.25$ &  $19.8$ &  $3.6$ &  $0.040$
      &  $\phantom{0}0.4$ &  $0.16$ &  $3.6$ &  $\phantom{0}1.5$ &  $0.104$ \\
  \hline
    \texttt{HY\_RP7.5}

      & $11.08$ & $-$ & $\phantom{0}7.02$ & $0.04$ & $13.4$ & $2.1$ & $3.681$
      & $\phantom{0}8.9$ & $0.81$ & $1.5$ & $16.8$ & $2.170$ \\
    \texttt{LK\_RP7.5\_LR}$^{a,c}$
      & $11.08$ & $-$ & $\phantom{0}2.31$ & $0.14$ & $10.6$ & $1.6$ & $0.687$
      & $\phantom{0}5.9$ & $0.45$ & $1.8$ & $15.8$ & $0.179$ \\
    \texttt{LK\_RP7.5}

      & $11.08$ & $-$ & $\phantom{0}2.67$ & $0.14$ & $10.6$ & $1.7$ & $0.890$
      & $\phantom{0}6.2$ & $0.49$ & $1.8$ & $15.5$ & $0.274$ \\
    \texttt{M0\_RP7.5}

      & $11.08$ &  $-$ &  $\phantom{0}4.55$ &  $0.16$ &  $10.5$ &  $1.8$ &  $1.656$
      &  $\phantom{0}7.8$ &  $0.62$ &  $1.6$ &  $17.2$ &  $0.609$ \\
  \hline
    \texttt{HY\_RP10}

      & $14.78$ & $-$ & $12.54$ & $0.04$ &  $\phantom{0}9.3$ & $1.5$ & $2.952$
      & $14.3$ & $0.78$ & $1.4$ & $29.0$ & $0.628$ \\
    \texttt{LK\_RP10}

      & $14.78$ & $-$ & $\phantom{0}5.37$ & $0.09$ & $\phantom{0}8.2$ & $1.7$ & $1.812$
      & $\phantom{0}8.6$ & $0.65$ & $1.6$ & $18.5$ & $0.621$ \\
    \texttt{M0\_RP10}

      & $14.78$ &  $-$ &  $\phantom{0}5.41$ &  $0.10$ &  $\phantom{0}8.3$ &  $1.7$ &  $1.829$
      &  $\phantom{0}8.6$ &   $0.65$ &   $1.6$ &   $18.5$ &   $0.629$ \\
  \hline
    \texttt{LK\_RP15}$^b$

      & $22.15$ & $-$ & $-$ & $-$ & $-$ & $-$ & $-$
      & $-$ & $-$ & $-$ & $-$ & $-$ \\
  \hline\hline
  \multicolumn{12}{l}{$^a$ This simulation uses the CMA scheme (see Sect.
  \ref{sec:grhydro}).} \\
  \multicolumn{12}{l}{$^b$ The ejected mass for this model is too small to be
  reliably measured.} \\
  \multicolumn{12}{l}{$^c$ This is a low-resolution simulation. Grid spacing on the
  inner AMR level of $h \simeq 369$~m.} \\
  \end{tabular}
  }
  \end{center}
\end{table*}

\noindent We consider two families of initial data. One describing two \acp{NS}
on Newtonian parabolic orbits with varying (Newtonian) periastron radius $r_p$,
and another family describing binaries in quasi-circular (low eccentricity)
orbits. In both cases, we fix the mass ratio to one (equal mass binaries). A
summary of all evolved models is presented in Tab. \ref{tab:models}.

We construct the eccentric initial data using binaries where each star has
gravitational mass at infinite separation $M_\infty = 1.389\ M_\odot$ and
baryonic mass $M_b = 1.522\ M_\odot$. We prepare the initial data by
superimposing two \ac{TOV} \citep{tolman:39, oppenheimer:39} solutions in
neutrino-less beta equilibrium at the initial separation of $100\ M_\odot \simeq
148\ \mathrm{km}$.  We set the orbital velocity of the two stars according to
Newton's law to have a Newtonian periastron radius of $0, 5\, M_\odot, 7.5\,
M_\odot, 10\, M_\odot$, and $15\, M_\odot$ ($\simeq 0, 7.38\, \mathrm{km},
11.08\, \mathrm{km}, 14.76\, \mathrm{km}$, and $22.15\,\mathrm{km}$).

Since this construction does not yield an exact solution to Einstein equations,
it results in violations of the Hamiltonian constraint equations that are
between $1$ and $2$ orders of magnitude larger than for quasi-circular binary
initial data. These values are within an acceptable range in that the
constraints remain well behaved during the evolution. On the other hand, since
we do not solve for the hydrostatic equilibrium of the \acp{NS} after boosting
them, this triggers oscillations in the two \acp{NS} with typical amplitudes
$\delta \rho_{\max}/\rho_{\max} \sim 0.15$. More accurate initial data could be
obtained using the methods described in \citet{east:12c} or
\citet{moldenhauer:14}. Alternatively, the errors in the constraint could be
mitigated adopting constraint-damping formulations of the Einstein equations
\citep{bernuzzi:10, weyhausen:11, alic:11a, hildich:13, alic:13, kastaun:13}.
Note, however, that none of these methods, with the exception of
\citet{moldenhauer:14}, solves for the hydrostatic equilibrium and can remove
all of the oscillations. We leave the investigation of these methods to future
work.

We construct the quasi-circular (low eccentricity) initial data using the
\texttt{LORENE} pseudo-spectral elliptic solver \citep{gourgoulhon:01}. We setup
irrotational initial data at a separation of $45\ \mathrm{km}$ consisting of
\acp{NS} each having gravitational mass at infinite separation $M_\infty =
1.384\ M_\odot$ and baryonic mass $M_b = 1.515\ M_\odot$.

For the evolution we make use of the \ac{AMR} capabilities provided by the
\texttt{Carpet} \citep{schnetter:03b} mesh refinement driver for \texttt{Cactus}
\citep{goodale:03}. During the inspiral, we employ a grid composed of 5
refinement levels with the finest ones composed of boxes that move to follow the
centers of the two \acp{NS}. After merger, we switch to a fixed grid, also
composed of 5 refinement levels. In both cases, the finest grid, which covers
the two neutron stars and the merger remnant, has a grid spacing of $h = 0.145\
M_\odot \simeq 215\ \mathrm{m}$. We also perform a simulation at lower
resolution with $h = 0.25\ M_\odot \simeq 369\ \mathrm{m}$ to estimate
finite-resolution effects on our simulations. To reduce the computational cost,
we exploit the symmetries of the problem to restrict our calculations to $x \geq
0$, $z \geq 0$: rotational symmetry is used across the $yz-$plane and reflection
symmetry is uses across the $xz-$plane.

\section{Results}
\label{sec:results}
\subsection{Overall Dynamics}
\label{sec:results.dynamics}
Eccentric and quasi-circular \ac{BNS} mergers have rich dynamics that can lead
to a number of outcomes depending on the \ac{EOS}, masses, and orbital
parameters of the binary \citep{shibata:06, baiotti:08, rezzolla:10, gold:12,
east:12b, paschalidis:15, east:15b}. This is also reflected in the dynamics of
our simulations. Binaries with small Newtonian periastron radii (\texttt{RP0} --
head-on collision -- and \texttt{RP5}) result in prompt \ac{BH} formation and
negligible amount of unbound mass. Binaries with larger Newtonian periastron
radii, \eg, \texttt{RP7.5}, and quasi-circular binaries (\texttt{QC}) result in
the formation of \acp{HMNS} \citep{baumgarte:00}, meta-stable massive remnants
temporarily supported against gravitational collapse by a large degree of
differential rotation \citep{baiotti:08, kaplan:14}. Binaries with
larger Newtonian periastron radii (models \texttt{RP10} and \texttt{RP15})
result in multiple close encounters.  The \texttt{RP10} model, in particular,
has a very complex dynamics and undergoes three close encounters, before finally
merging after the third encounter. The \texttt{RP15} model is also expected to
undergo multiple encounters, however the timescales between successive
encounters is too long to be simulated with our methods, so we only consider its
first close passage.

The orbital dynamics of the \texttt{RP10} model appears to be very sensitive to
small changes in the simulation inputs and, in our preliminary tests, we
observed differences in the timing and number of encounters with changes in the
numerical parameters (\eg, resolution). A similar behavior is also observed in
eccentric binary \ac{BH}s encounters \citep{damour:14, guercilena:15}. For this
reason, the \texttt{RP10} simulations should be considered as particular
realizations of the dynamics close to the threshold between direct merger and
multiple encounters and not necessarily as the outcomes of encounters with $r_p$
exactly equal to $10\ M_\odot$. Indeed, the precise \emph{value} of this
threshold is probably the imprint of our numerical setup and is likely to change
once this setup is varied. On the other hand, the \emph{existence} of such a
threshold will not depend on the numerical details.  The orbital dynamics of the
other binaries, instead, appears robust.  The accuracy of numerical relativity
simulations of quasi-circular \ac{BNS} inspirals has been studied in detail in
the past (\mbox{\citealt{baiotti:09};} \citealt{baiotti:10, bernuzzi:12,
radice:14a, radice:14b, radice:15b}).

\begin{figure*}
  \includegraphics[width=\textwidth]{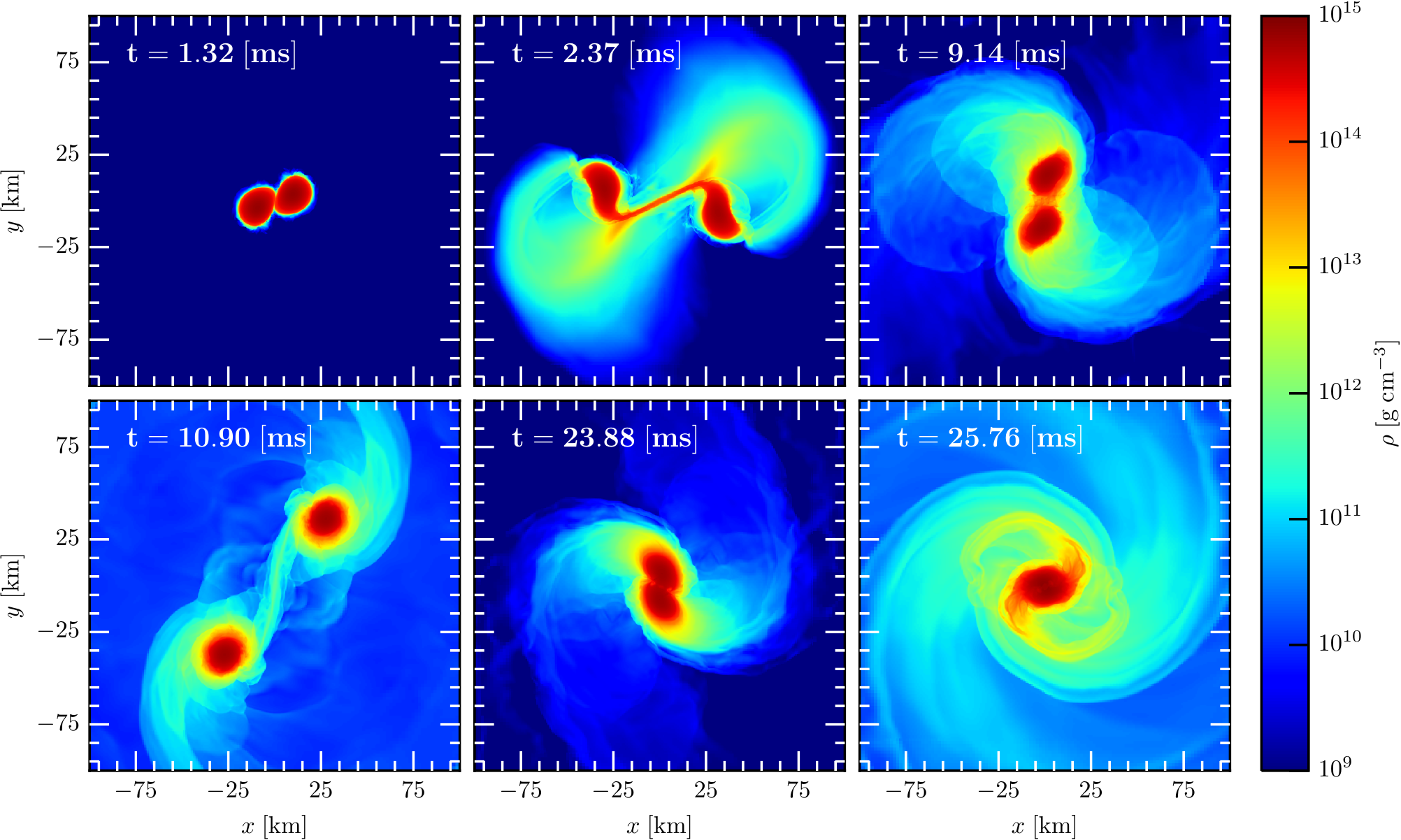}
  \caption{Rest-mass density (equation \ref{eq:rmd}) in the orbital plane for
  the parabolic encounter simulation \texttt{LK\_RP10} at six different times.
  The \ac{NS}s undergo three close encounters before merging. The panels show
  snapshots of the two stars immediately before and after each encounter. Tidal
  torques at the periastron result in large mass ejection and trigger
  oscillations in the \acp{NS}, \cf\ Fig. \ref{fig:psi4}.}
  \label{fig:rho_xy}
\end{figure*}

The dynamics of binary \texttt{LK\_RP10} is shown in Fig. \ref{fig:rho_xy}.
There we plot color maps of the rest-mass density in the orbital plane at
representative times during the evolution (before and after each close
encounter). During the periastron passage strong tidal torques and shocks result
in episodic outflow events. Part of the ejected neutron-rich matter is unbound
from the system (more on this in Sect. \ref{sec:results.ejecta}), while the rest
settles in a thick atmosphere around the \ac{NS}s. The atmosphere is mostly
thermally supported in the purely hydrodynamic \texttt{HY\_RP10} simulation, but
not in the simulations which include neutrino cooling. In the latter cases, the
``envelope'' around the binary has time to cool and partly accrete back onto the
\ac{NS}s in the time between successive encounters.  This results in the
\texttt{LK\_RP10} simulation undergoing its last encounter $\sim 2\ \mathrm{ms}$
before \texttt{HY\_RP10}.

\begin{figure}
  \includegraphics[width=\columnwidth]{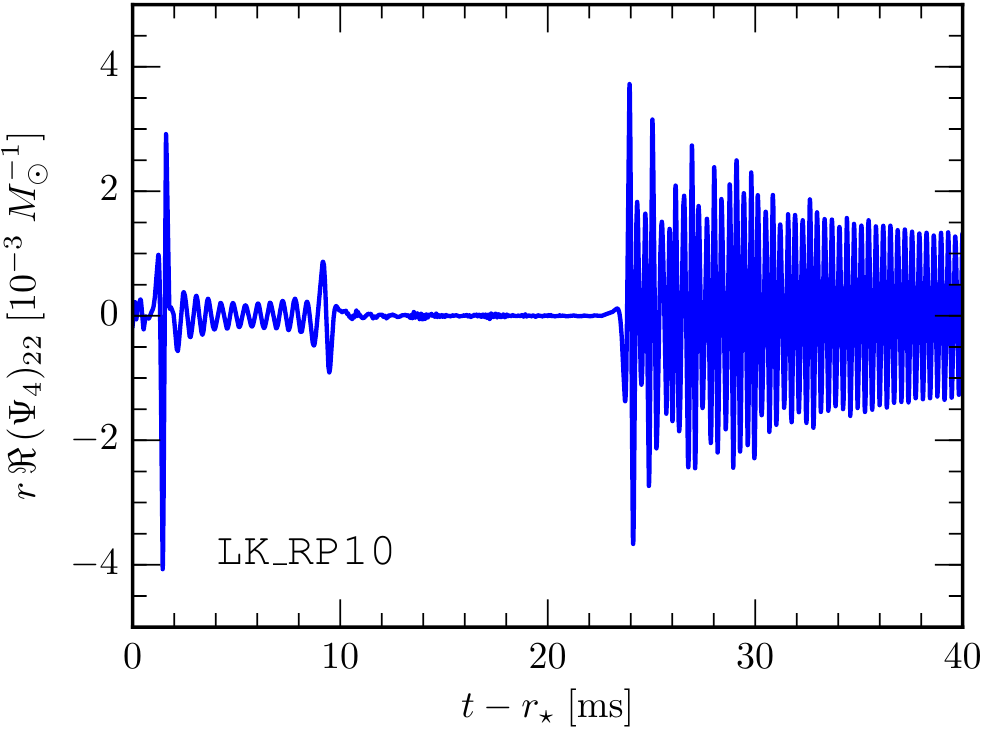}
  \caption{Real part of the $\ell = 2,\ m = 2$ spin-weighted spherical harmonics
  component of the Weyl scalar $\Psi_4$ for the \texttt{LK\_RP10}, extracted at
  $r_\star = 400\ M_\odot \simeq 590\ \mathrm{km}$. The curvature \ac{GW} signal
  shows a burst after the first encounter that excites violent oscillations in
  the two \ac{NS}s. These oscillations are then suppressed by tidal interactions
  during the second encounter. The \ac{GW} signal turns on again at merger.}
  \label{fig:psi4}
\end{figure}

The encounters also excite oscillations of the two \acp{NS}, which result in
copious \ac{GW} emission. This phenomenon has been previously reported in
simulations employing idealized \ac{EOS} \citep{gold:12, east:12b} and studied
in detail by \citet{gold:12}. They showed that tidal interactions during close
encounters can excite the fundamental modes of oscillation of the two \acp{NS}.
This is apparent in our \texttt{RP10} model whose curvature \ac{GW} signal, the
Weyl scalar $\Psi_4$\footnote{We remind the reader that the complex scalar
$\Psi_4$ combines the second time derivatives of the two strain polarizations
\mbox{$\Psi_4 = \ddot{h}_+ - \mathrm{i}\, \ddot{h}_\times$}.}, is shown in Fig.
\ref{fig:psi4}. The first encounter is accompanied by a burst in \acp{GW}
followed by a quasi-periodic signal lasting about $\sim 10\ \mathrm{ms}$, \ie,
up to the time of the second encounter. For all of the \texttt{RP10}
simulations, we find the interaction between the two \acp{NS} during the second
encounter to suppress the oscillations of the two stars leading to a sudden
shutdown of the \ac{GW} emission. This is an effect that has not been reported
before. However, it may be a consequence of the $\pi-$symmetry imposed during
the evolution.  More simulations would be required to address this question. A
second, sudden, burst in \acp{GW} appears at the time of merger.

Like the orbital dynamics of the binaries before the merger, also their
post-merger evolution shows a large variety. As discussed above, the merger can
result in prompt \ac{BH} formation for binaries with little angular momentum
(\texttt{RP0}, \texttt{RP5}) or in the formation of a \ac{HMNS}.  Furthermore,
since different models merge with different amount of residual angular momentum
and mass, the structure and rotational configuration of the \ac{HMNS} can vary
significantly leading to different evolutionary paths.

\begin{figure}
  \includegraphics[width=\columnwidth]{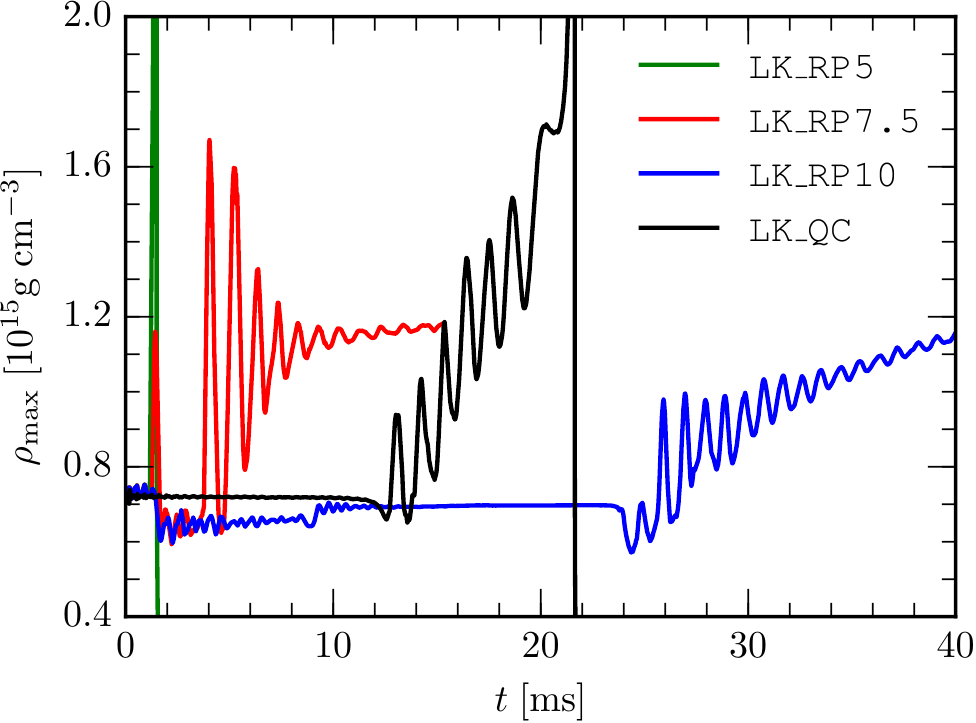}
  \caption{Maximum rest-mass density (equation \ref{eq:rmd}) as a function of
  time for the simulations with neutrino cooling, \texttt{LK} runs. The
  \texttt{LK\_RP} runs are with eccentric binaries, while the \texttt{LK\_QC} is
  quasi-circular. Although all of the models have almost the same mass and we
  use a single \ac{EOS}, there is significant variability in the outcome of the
  merger due to the differences in the amount of angular momentum of the binary
  at merger.}
  \label{fig:rhoc}
\end{figure}

This is summarized in Fig. \ref{fig:rhoc}, where we show the maximum rest-mass
density as a function of time for a group of representative simulations. The
\texttt{RP5} model exhibits prompt collapse: the central density rapidly grows
until an apparent horizon is formed. After the formation of the apparent
horizon, $\rho_{\max}$ appears to drop, because we exclude the region inside the
\ac{BH} in the calculation of the maximum density. The \texttt{RP7.5} model
experiences a rather violent merger: the two stellar cores collide, slightly
overshoot each other, and then merge again after $\sim 2\ \mathrm{ms}$. In
comparison, the quasi-circular model \texttt{QC} experiences a much milder
merger, with a smaller initial jump in the density. However, since the
\texttt{QC} binaries merge with smaller residual total angular momentum than the
\texttt{RP7.5} binaries, the resulting \ac{HMNS} contracts and oscillates
violently before collapsing to a \ac{BH} within $\sim 10\ \mathrm{ms}$ after
merger. Finally, the \texttt{RP10} binary merges with a large amount of residual
angular momentum, which sustains the hypermassive merger remnant over the entire
simulated time ($\gtrsim 20\ \mathrm{ms}$ after merger).

We report the remnant torus masses for those binaries resulting in \ac{BH}
formation within the simulated time in Tab.~\ref{tab:models}. The torus masses
are quoted at $1\ \mathrm{ms}$ after the formation of the apparent horizon. We
find very small torus masses ($< 10^{-2}\ M_\odot$) for the nearly head-on
eccentric mergers, which result in prompt \ac{BH} formation. Larger torus masses
($\sim \textrm{few} \times 10^{-2}\ M_\odot$) are found for the \texttt{QC}
models. The latter, however, vary by more than a factor of two depending on the
level of microphysical description.  The purely hydrodynamical simulation
produces the largest torus mass and the simulation including neutrino cooling,
but not heating, produces the smallest.

\begin{figure*}
  \includegraphics[width=\textwidth]{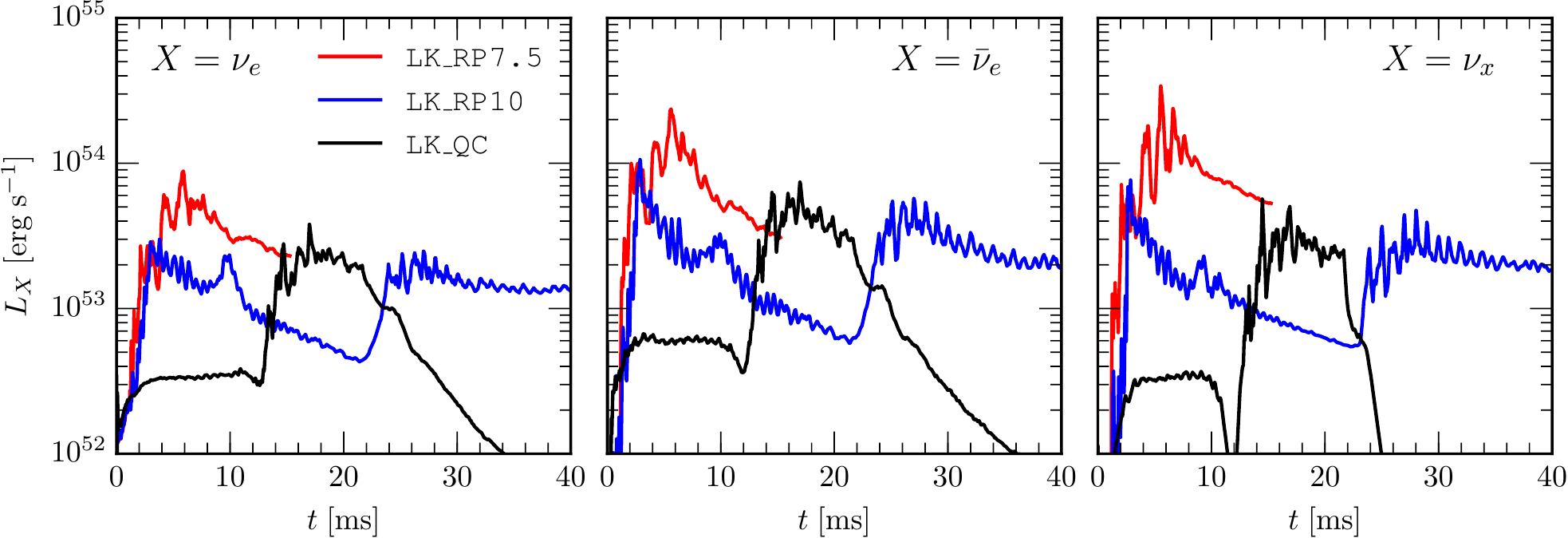}
  \caption{Neutrino luminosity for runs \texttt{LK\_RP7.5}, \texttt{LK\_RP10},
  and \texttt{LK\_QC}. \emph{Left panel:} electron neutrino luminosity.
  \emph{Middle panel:} electron anti-neutrino luminosity. \emph{Right panel:}
  total luminosity from heavy lepton ($\mu, \bar{\mu}, \tau, \bar{\tau}$)
  neutrinos.  Models with small periastron radius, such as \texttt{LK\_RP7.5},
  result in violent mergers with very bright neutrino bursts.}
  \label{fig:luminosity}
\end{figure*}

The diversity in the merger dynamics is also reflected in the neutrino signals
shown in Fig. \ref{fig:luminosity}. The violent merger in the
\texttt{LK\_RP7.5} simulation results in neutrino luminosities in excess of
$10^{54}\ \mathrm{erg}\ \mathrm{s}^{-1}$ sustained over multiple milliseconds.
These luminosities are comparable to those reported by \citet{rosswog:13}, who,
however, performed simulations using Newtonian gravity and a different nuclear
\ac{EOS}. These luminosities are almost an order of magnitude larger than for
our fiducial quasi-circular simulation (model \texttt{QC\_LK}). The
\texttt{LK\_RP10} simulation displays sudden rises in its neutrino emissions in
coincidence with each close encounter, followed by cooling phases where the
luminosity drops exponentially in time. Finally, the \texttt{QC\_LK} run shows a
single burst in its neutrino emission at the time of merger, followed by a rapid
decay in its luminosity as soon as the \ac{HMNS} collapses to a \ac{BH}. The
early time, $t \lesssim 12\ \mathrm{ms}$, neutrino emissions from the
\texttt{QC\_LK} simulation are due to the spurious heating of the surface of the
two \acp{NS} caused by our numerical scheme. These early-time neutrino energy
losses might appear large, but they are actually not dynamically important as
they are $\sim 2$ orders of magnitude smaller than the \ac{GW} luminosity, which
peaks at $\sim 2\times 10^{55}\ \mathrm{erg}\ \mathrm{s}^{-1}$ at the time of
merger.

\subsection{Dynamical Ejecta}
\label{sec:results.ejecta}
\subsubsection{Ejected Mass}

\begin{figure}
  \includegraphics[width=\columnwidth]{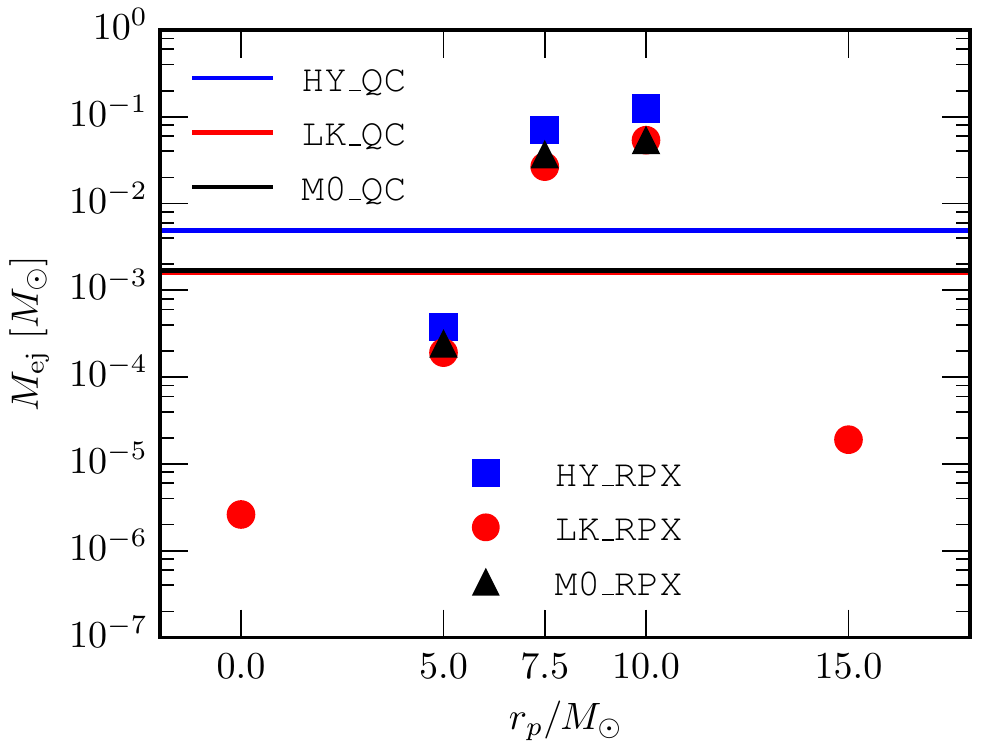}
  \caption{Dynamically ejected mass for all simulations as a function of the
  Newtonian periastron radius $r_p$. The ejecta mass is computed integrating in
  time the flux of unbound matter (with $u_t \leq -1$) across the surface of a
  spherical sphere with radius $r = 200\ M_\odot \simeq 295\ \mathrm{km}$. The
  horizontal lines denote the ejected mass from the quasi-circular runs. The
  ejecta for models with $r_p = 0$ and $r_p = 15\ M_\odot \simeq 22.15\
  \mathrm{km}$ are under-resolved and should not be taken at face value.
  Overall, eccentric binaries can eject up to $2$ orders of magnitude more mass
  than quasi-circular binaries.}
  \label{fig:ejecta.mass}
\end{figure}

\noindent Being on Newtonian parabolic orbits, our eccentric \ac{BNS} binaries
are only weakly bound. For this reason, it is not surprising to find that they
can unbind a significantly larger amount of matter compared to quasi-circular
mergers.  Figure \ref{fig:ejecta.mass} shows the total amount of unbound matter
for each of our simulations (also reported in Tab. \ref{tab:models}). The total
ejected rest mass is computed integrating in time the flux of matter with $u_t
\leq -1$\footnote{See \citet{kastaun:15} for a discussion of possible
alternative criteria to identify unbound fluid elements.} across a spherical
coordinate surface with radius $200\ M_\odot \simeq 295\ \mathrm{km}$.  We find
that this choice can result in an underestimate of the total ejecta mass by up
to $\sim 20\%$ as more material can become unbound at larger radii. However, the
data extracted from larger radii is potentially affected by unphysical
artifacts, because the density drops to values closer to the floor and the
assumption of nuclear statistical equilibrium (assumed in our \ac{EOS}
treatment) is violated.  Also note that these estimates do not include late-time
mass ejection driven by neutrino, viscous heating, and/or magnetic pressure,
which would take place on longer timescales \citep{dessart:09, fernandez:13,
siegel:14, metzger:14, fernandez:14, rezzolla:14, ciolfi:14, fernandez:15,
just:15a, martin:15, kiuchi:15a}. 

\begin{figure}
  \includegraphics[width=\columnwidth]{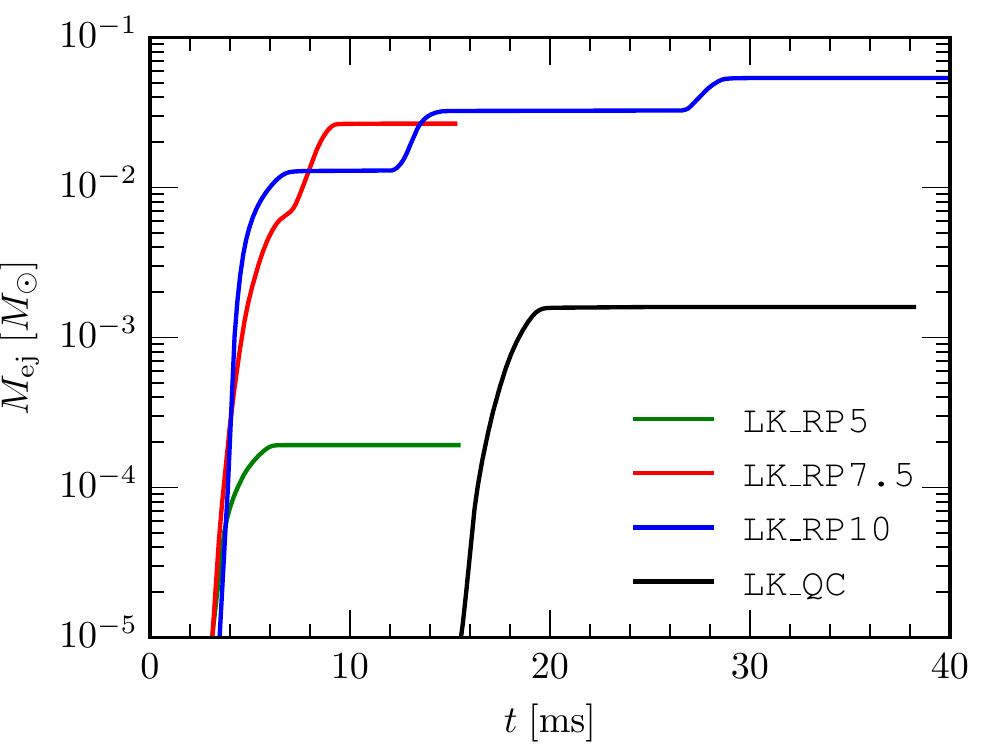}
  \caption{Total ejected mass for the \texttt{LK\_RP5}, \texttt{LK\_RP7.5},
  \texttt{LK\_RP10}, and \texttt{LK\_QC} models. The ejecta mass is computed
  by integrating the flux of unbound matter (with $u_t \leq -1$) across a
  coordinate-sphere with radius $r = 200\ M_\odot \simeq 295\ \mathrm{km}$. The
  dynamical mass ejection is impulsive and lasts only a few milliseconds.}
  \label{fig:outflow}
\end{figure}

As shown in Fig.~\ref{fig:outflow}, the dynamical mass ejection in our
simulations is impulsive and concentrated in one or, in the \texttt{RP10}
models, three ejection events lasting only several milliseconds. For this
reason, our measure of the dynamically ejected mass is robust with respect to
the physical time covered by our simulations.

Some interesting trends can be observed from Fig. \ref{fig:ejecta.mass}.  First
of all, we find that eccentric mergers, and in particular mergers with impact
parameters close to the threshold between prompt merger and multiple-encounters,
seem to be those resulting in the largest mass ejection.  This is similar to
what was found by \citet{east:12b}. Note that the \texttt{RP15} model is likely
to experience more mass ejection in its successive encounters, which will happen
over timescales that we cannot simulate directly.  Nevertheless, we still expect
encounters such as that of the \texttt{LK\_RP15} run to yield lower ejecta mass
compared to encounters resulting in mergers over a short dynamical timescale,
such as \texttt{RP7.5} and \texttt{RP10}. The reason for this is that model
\texttt{RP15} will undergo merger in a more gravitationally bound state than
model \texttt{RP7.5} or model \texttt{RP10}. As a consequence, we conjecture on
the basis of our data that eccentric \ac{BNS} mergers can yield up to $\sim 0.1\
M_\odot$ of ejecta, slightly less than what can be achieved with \ac{BHNS}
mergers \citep{foucart:14, kyutoku:15}, but almost two orders of magnitude
larger than what is ejected by mergers of BNSs in quasi-circular orbits.

We find significant differences between the \texttt{LK} results, which include
neutrino cooling, and the \texttt{HY}, which neglect it. For all our models we
find that neglecting neutrino cooling results in an overestimate of the unbound
mass by a factor $\gtrsim 2$. This hints at the importance of neutrino-radiation
processes in shaping the outflows from these mergers. As documented in detail by
\citet{hotokezaka:13} and \citet{bauswein:13}, a significant fraction of the
ejecta in \ac{GR} simulations is driven by shocks.  In our simulations neutrino
losses in the optically thin outflows are rapid and sufficient to cause part of
the material to become gravitationally bound again by removing part of its
internal energy. In addition, the total amount of mass ejected by the
\texttt{RP10} binaries is also affected by the cooling of the thick atmosphere
generated during the first encounter of the two \acp{NS}, which is suppressed in
the \texttt{HY\_RP10} run, as discussed in Sect.~\ref{sec:results.dynamics}.

The differences between models including only cooling (\texttt{LK}) and models
also including heating (\texttt{M0}) are also important, but less striking. We
find that the inclusion of heating results in
a slight increase of the total ejected mass,similarly to what has been
reported by \cite{sekiguchi:15}.

\subsubsection{Properties of the Outflow}

\begin{figure*}
  \begin{center}
    \includegraphics[width=2\columnwidth]{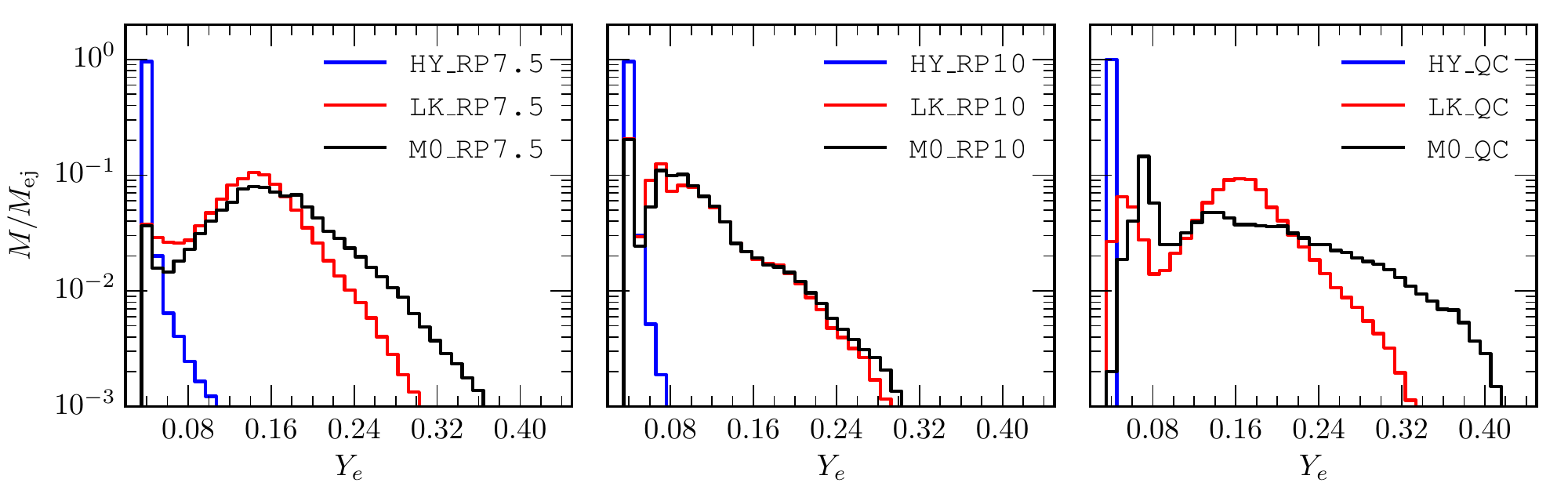}\\[1em]
    \includegraphics[width=2\columnwidth]{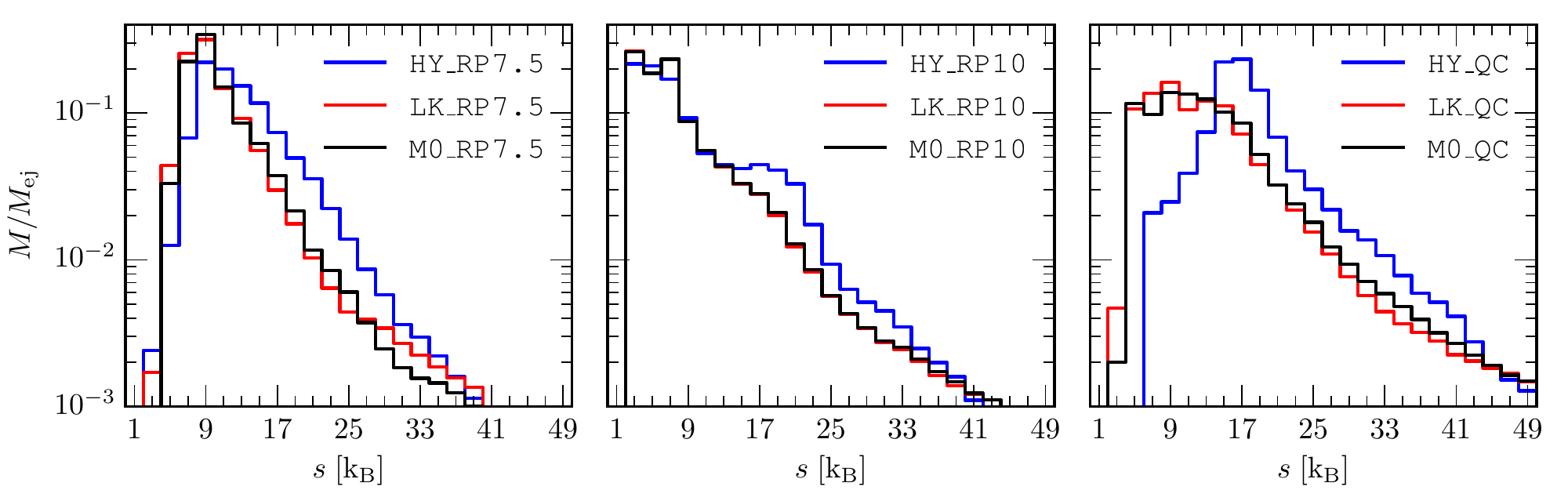}\\[1em]
    \includegraphics[width=2\columnwidth]{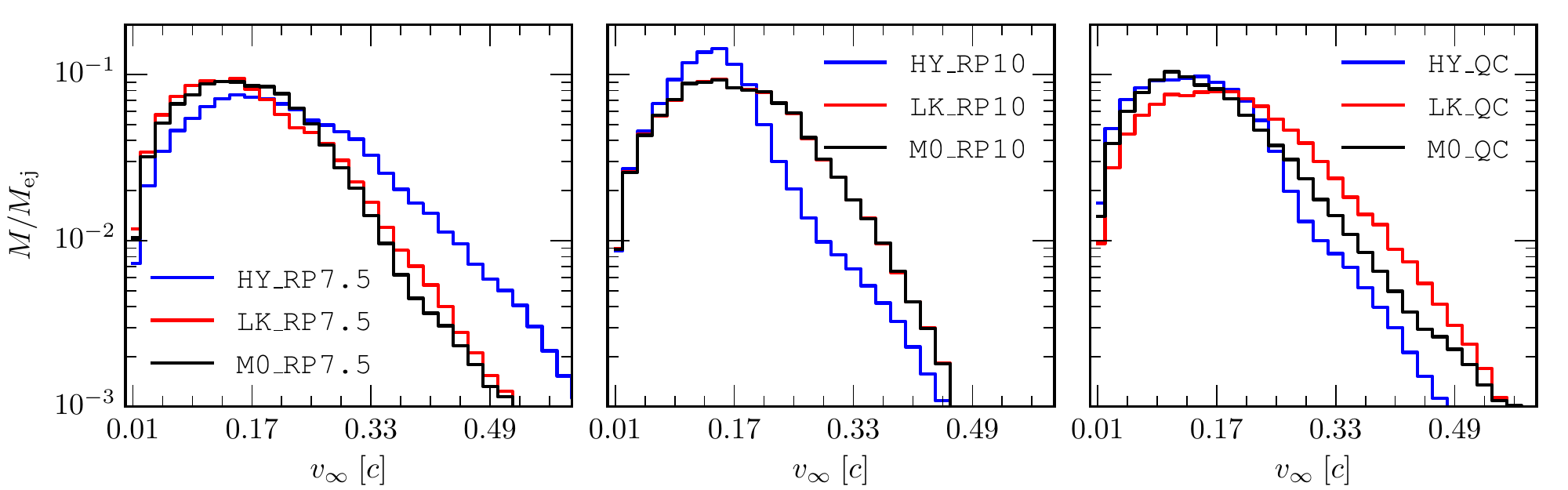}\\[1em]
    \includegraphics[width=2\columnwidth]{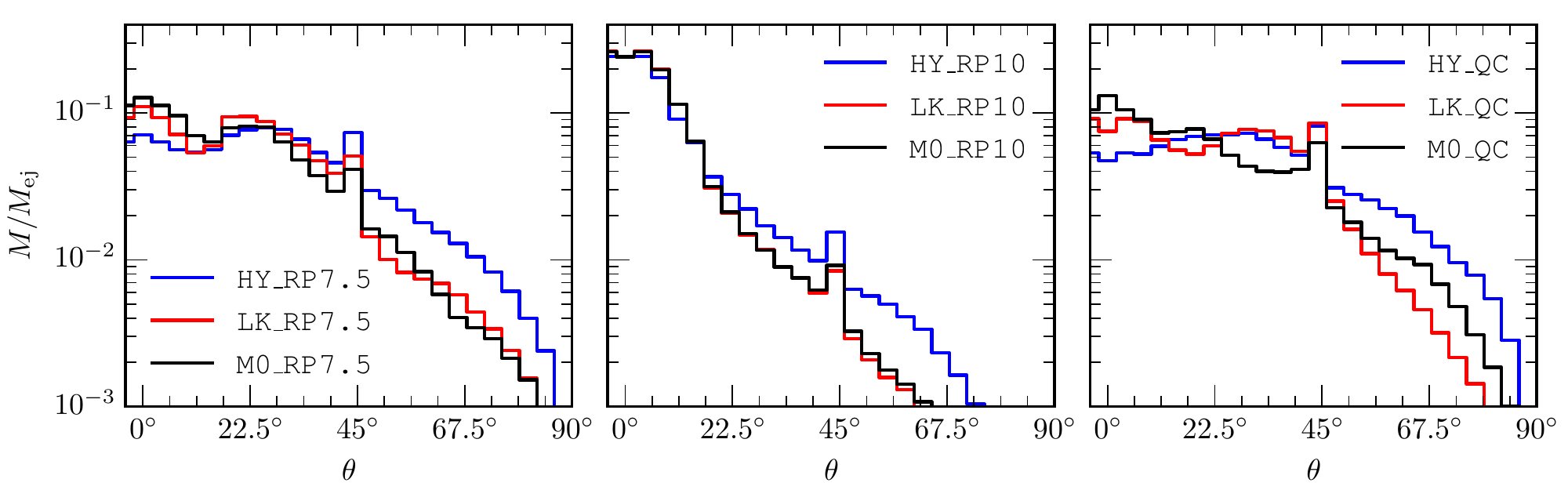}
  \end{center}
  \caption{Electron fraction (\emph{top row}), specific entropy per baryon
  (\emph{second row}), asymptotic velocity (\emph{third row}), and angular
  distribution (\emph{bottom row}) of the ejecta. $\theta$ is the angle from the
  orbital plane. The first, second, and third columns show results from models
  \texttt{RP7.5}, \texttt{RP10}, and \texttt{QC} respectively. For each
  configuration we consider three different levels of microphysical description:
  pure hydrodynamics (\texttt{HY}), neutrino cooling (\texttt{LK}), or neutrino
  cooling and heating (\texttt{M0}). The histograms are computed from the mass
  fraction of the matter crossing a spherical surface at radius $r = 200\
  M_\odot \simeq 295\ \mathrm{km}$ with positive specific energy (\ie, with $u_t
  \leq -1$). The bump in the angular distribution at $\theta = 45^\circ$ is a
  numerical artefact generated by our Cartesian simulation grid.}
  \label{fig:histograms}
\end{figure*}

\noindent We collect composition, specific entropy, asymptotic velocity and
angular distribution of the ejecta for the three most interesting models,
\texttt{RP7.5} (representative of a violent collision), \texttt{RP10}
(representative of the dynamics with multiple encounters), and \texttt{QC} (our
fiducial quasi-circular binary) in Fig. \ref{fig:histograms}. The histograms are
generated by binning the properties of the unbound matter (with $u_t \leq -1$)
flowing through a coordinate sphere with radius $r = 200\ M_\odot \simeq 295\
\mathrm{km}$. As can be seen from Fig. \ref{fig:histograms}, morphology and
thermodynamical properties of the ejecta show large variations with binary
configuration and neutrino treatment. This is a consequence of the complex
interplay, between radiation and hydrodynamics, that controls the mass ejection.

The electron fraction of the ejecta is one of the most important quantities
determining both the outcome of the nucleosynthesis (more in Sect.
\ref{sec:results.nucleosynthesis}) and the properties of the \ac{EM} signature
of the merger (see Sect. \ref{sec:results.EM}). It is also the quantity showing
the greatest variation. Simulations neglecting all weak interactions are,
unsurprisingly, characterized by extremely neutron-rich ejecta. In these
simulations, $Y_e$ is simply advected with the flow and unchanged. On the other
hand, when weak interactions are included, $Y_e$ can evolve due to electron or
positron captures in shock-heated material (in the \texttt{LK} and \texttt{M0}
simulations) and due to neutrino absorption (in the \texttt{M0} simulations). As
a consequence, simulations accounting for weak interactions show a much wider
range of electron fractions.

Independent of binary parameters, we find the ejecta to have a clear bi-modal
distribution in $Y_e$. Part of the outflow, driven by tidal torques, is cold and
neutron rich, while another component, launched by shocks during merger,
experiences high-temperatures and rapid protonization with values of $Y_e$
peaking at $\sim 0.16$. The \texttt{M0} simulations also show a relatively
proton-rich $Y_e \simeq 0.4$ component of the outflow, predominantly at high
latitudes (more on this later). The final mass distribution of $Y_e$ depends on
the relative importance of these different components.

The composition of the ejecta from our simulations is somewhat more neutron rich
than that of \citet{wanajo:14} and \citet{sekiguchi:15}. These studies appear to
be lacking any cold component in their outflows. This might be a consequence of
the different treatment of the neutrino radiation or of the choice of \ac{EOS}.
Instead, we find good qualitative agreement between our \texttt{LK\_QC} results
and those of \citet{palenzuela:15}, who adopted different \ac{EOS}, but a
neutrino cooling treatment very similar to ours. Our ejecta are also
significantly more proton-rich than those typically found in Newtonian
simulations \citep[\eg,][]{rosswog:13}, where the tidal component is enhanced
with respect to \ac{GR} simulations \citep{bauswein:13}.

The composition of the ejecta differs between eccentric and quasi-circular
mergers. The reason for this is that eccentric mergers producing large ejecta
mass do so mostly as a consequence of the enhanced tidal interaction between the
two \ac{NS}s during close passages. As a consequence, only nearly head-on
collisions (model \texttt{RP5}) result in more proton-rich ejecta than the
\texttt{QC} binaries (see Tab. \ref{tab:models}). This trend is the
\emph{opposite} of what has been reported in Newtonian simulations, where
quasi-circular binaries are dominated by tidal ejecta with low-$Y_e$ and
eccentric binaries by shock-heated, high-$Y_e$, ejecta \citep{rosswog:13}. The
\texttt{LK\_RP7.5} simulation has ejecta composition that is very similar to
that of the quasi-circular \texttt{LK\_QC}, while the multiple encounter
simulation \texttt{LK\_RP10} has significantly more neutron rich ejecta than
both.

The effect of neutrino absorption on the electron fraction is non-negligible, as
also found by \citet{wanajo:14}, \citet{sekiguchi:15}, and \citet{foucart:15b},
although in our simulations the impact is less significant than in these
previous studies. The \texttt{M0} runs have more proton rich ejecta than the
\texttt{LK} simulations, but the differences appear to be mostly confined to the
tail of the ejecta distribution in $Y_e$, while the mass-weighted average of the
electron fraction $\langle Y_e \rangle$ is not strongly affected (\cf, Tab.
\ref{tab:models} and Fig.  \ref{fig:histograms}).

As can be seen from the second row of Fig. \ref{fig:histograms}, the specific
entropy per baryon of the ejecta shows somewhat smaller differences between
simulations than $Y_e$. The overall trend is that runs that include neutrino
cooling effects (\texttt{LK} and \texttt{M0}) display lower specific entropies
than purely hydrodynamical models, which overestimate the final entropy of the
shock-heated ejecta.

We find the bulk of the outflow to be sub-relativistic with asymptotic
velocities $v_\infty \lesssim c/3$ (third row of Fig. \ref{fig:histograms}) for
both eccentric and quasi-circular models. This is in contrast with Newtonian
simulations \citep{rosswog:13}, but is similar to other \ac{GR} studies
\citep{east:12b, sekiguchi:15, east:15b}. There are some differences in the
asymptotic velocities between \texttt{HY}, \texttt{LK}, and \texttt{M0}
simulations, but there is no clear trend.

We show the angular distribution of the ejecta in the bottom row of panels in
Fig. \ref{fig:histograms}. We find the outflow of the \texttt{RP10} model, which
ejects matter mostly due to tidal torques during its multiple encounters, to be
mostly contained in a narrow angle around the orbital plane. This is similar
what found for the ejecta of \ac{BHNS} mergers \citep{foucart:14, kyutoku:15}.
The other models show a larger angular spread of the outflow with the bulk of
the ejecta appearing to be contained within an angle $\theta \lesssim 60^\circ$
of the orbital plane.  The bump in the ejecta distribution function at $\theta =
45^\circ$ is a numerical artefact associated with our Cartesian simulation grid
that tends to funnel flows along its symmetry directions.

The amount of ejected material at high latitudes ($\theta \gtrsim 60^\circ$) is
also strongly dependent on the included physics in our simulations. Runs that do
not include neutrino cooling tend to overestimate, by a factor of a few, the
amount of matter in the polar region of the post-merger remnant. Conversely,
simulations that include neutrino cooling, but not heating, underestimate the
baryon contamination of the poles by a similar factor. These differences might
have an impact for the simulation of \ac{SGRB} engines, because baryon
contamination of the polar regions might prevent the launch of relativistic jets
in some \ac{SGRB} models \citep[\eg,][]{just:15b}.

\subsubsection{Ejection Mechanisms}

\begin{figure*}
  \includegraphics[width=\textwidth]{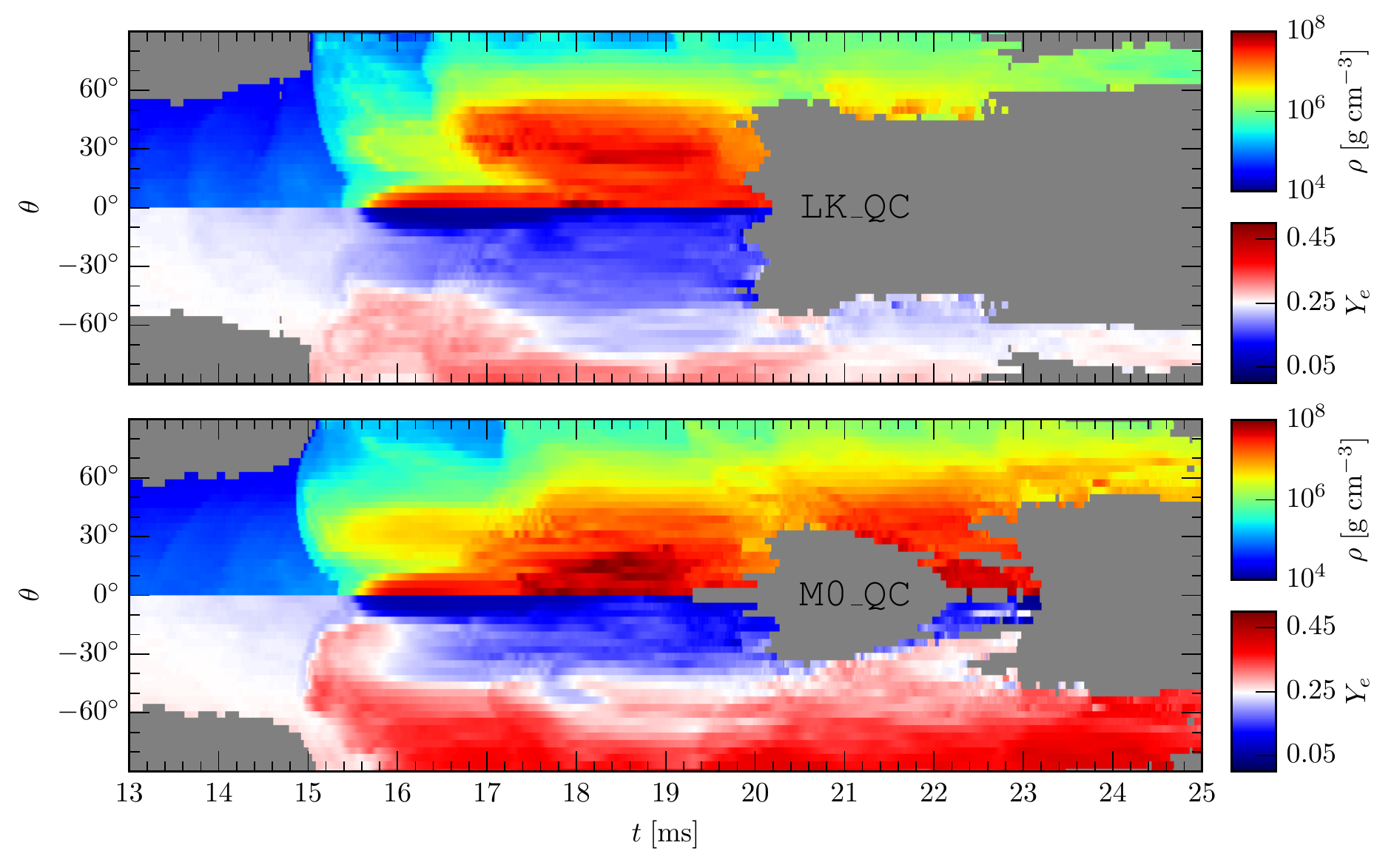}
  \caption{Angular distribution (\emph{upper half of each panel}) and
  composition (\emph{lower half of each panel}) of the ejecta for the
  \texttt{LK\_QC} (\emph{upper panel}) and \texttt{M0\_QC} (\emph{lower panel})
  simulations as a function of time. The data is collected on a coordinate
  sphere at radius $r = 200\ M_\odot \simeq 295\ \mathrm{km}$ and only considers
  the unbound part of the outflow (\ie, with $u_t \leq -1$). The gray shaded
  areas refer to times/angles for which we do not measure any outflow of unbound
  matter (\ie, where $u_t > -1$). The ejection event is of very short duration
  and the outflow is confined within a broad $\sim 60^\circ$ angle from the
  equator.  The material at low altitudes is typically more neutron rich than at
  higher altitudes, suggesting a different ejection mechanisms for the different
  components of the outflow.}
  \label{fig:ejecta.profile}
\end{figure*}

\noindent In Fig. \ref{fig:ejecta.profile}, we show the angular density and
$Y_e$ distribution of the ejecta for two of the quasi-circular \ac{BNS}
simulations: \texttt{LK\_QC} and \texttt{M0\_QC}. These profiles hint at the
interplay between a few distinct mechanism for the dynamical ejection of \ac{NS}
matter. A first component of the outflow is driven by tidal interactions between
the \acp{NS}. It is very neutron rich, with $Y_e \lesssim 0.1$, and confined in
a narrow $\sim 20^\circ$ angle from the orbital plane. In our simulations, this
is the first component to reach the fiducial outflow surface at $t \simeq 16\
\mathrm{ms}$ (\ie, $\simeq 3\ \mathrm{ms}$ after the merger; consistent with an
outflow velocity of $\sim c/3$). This component is present in all of our
simulations, regardless of their level of microphysical description.

A second component is driven by shocks launched during merger. It is less
neutron rich, with $Y_e \sim 0.15$, and more isotropic, spanning an angular
region of about $60^\circ$ from the orbital plane. Shock driven ejecta lag the
tidal tails by a few ms.

Finally, a third component is constituted by a high-latitude wind driven by a
combination of shock and neutrino heating. This component is relatively proton
rich with $Y_e \gtrsim 0.25$. It is mostly absent in the \texttt{LK\_QC}
simulation, but becomes the dominant outflow component after $t \simeq 21\
\mathrm{ms}$ in the \texttt{M0\_QC} run. The thermal wind also entrains a small
part of the accretion torus resulting in additional low-$Y_e$ material being
ejected close to the equatorial plane at late times.

The interplay between low-$Y_e$ (close to the equator) and high-$Y_e$ (at high
latitudes) ejecta is what determines the final electron fraction distribution of
the ejecta (Fig. \ref{fig:histograms}) and the r-process yields (more in Sect.
\ref{sec:results.nucleosynthesis}).

\subsection{Nucleosynthesis}
\label{sec:results.nucleosynthesis}
As the material ejected during merger expands, it undergoes neutron capture
nucleosynthesis, which may produce r-process elements.  In order to estimate the
final nucleosynthetic yields of our simulations, we map our ejecta data onto the
grid of parametrized r-process calculations presented in \citet{lippuner:15}.
First, we extract $Y_e$, $s$, $v$, and $\rho$ for all of the unbound matter
(with $u_t \leq -1$) as it flows across the surface of a coordinate sphere of
radius $r_E = 200\ M_\odot \simeq 295\ \mathrm{km}$.  To estimate the
nucleosynthesis in the outflow using our nucleosynthesis grid, the dynamical
timescale, $\tau$, of each ejected fluid element is required.  To estimate
$\tau$, we assume that the material crossing our extraction surface is expanding
homologously.  This assumption should be relatively well satisfied by our data
given that, by the time the ejecta reach the detection surface, their density
has already dropped by $\sim 3$ orders of magnitude.  Then, the density history
for a particular element of the ejecta is given by $\rho(t) = \rho_E (v_E
t/r_E)^{-3}$, where $\rho_E$ and $v_E$ are the density and velocity of the fluid
element when it crosses a sphere of radius $r_E$.  At late times, the density
history used in \citet{lippuner:15} has the form $\rho(t)=\rho(s, Y_e, T=6 \,
\textrm{GK}) (3\tau/et)^3$, where $e$ is Euler's number.  We match these two
profiles to extract $\tau$. The final yields are then found by multiplying the
yields computed by \citet{lippuner:15} at each point in their nucleosynthesis
grid with the total mass associated with it. Such a procedure is necessarily
approximate, but it enables the rapid calculation of the abundance distribution
in the ejecta without the need for tracer particles. Future work should
directly compare this method of yield estimation to more detailed
tracer-particle based methods.

We do not find material with expansion timescale of less than $0.5\
\mathrm{ms}$. This seems to exclude the neutron freeze-out scenario proposed by
\citet{metzger:15}. However, the lack of a very fast component of the ejecta
might also be due to numerical effects. Our resolution is probably not high
enough to track the very small fraction of the ejecta expected to experience
neutron freeze-out in the scenario proposed by \citet{metzger:15}.

\begin{figure*}
  \includegraphics[width=\textwidth]{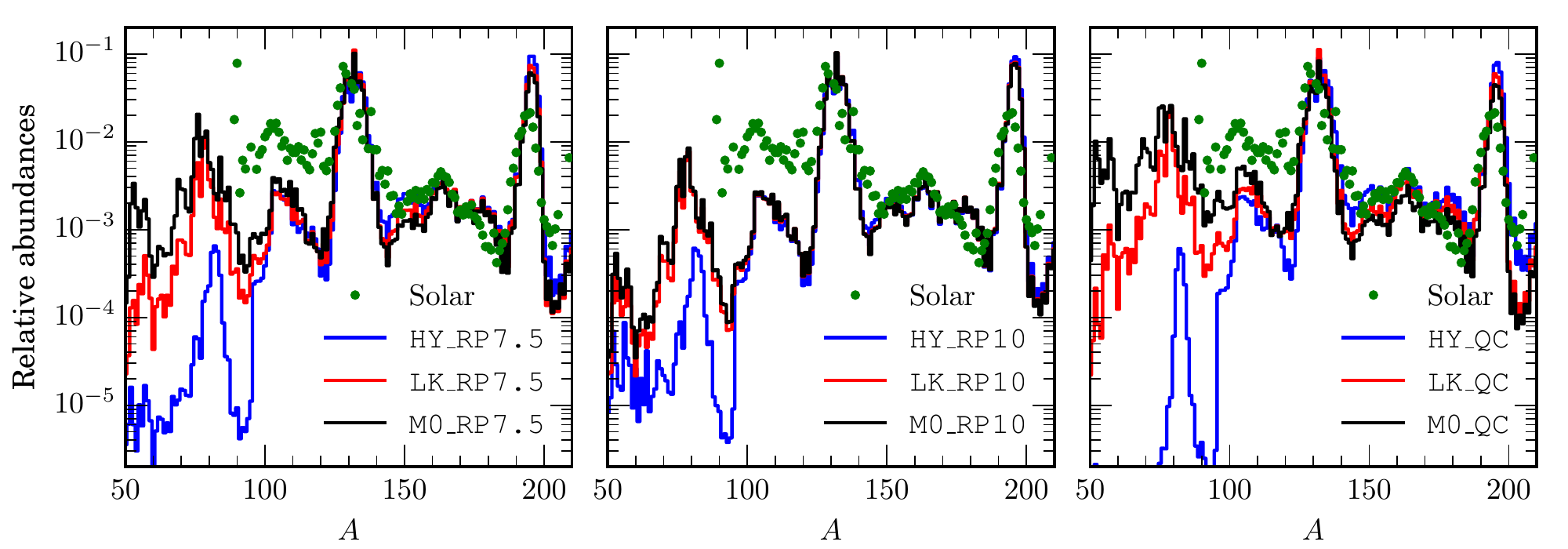}
  \caption{Final abundances in the ejecta for the \texttt{RP7.5}, \texttt{RP10}
  and \texttt{QC} configurations. The yields are normalized with the total
  abundance of elements with $63 \leq A \leq 209$. For each configuration we
  consider three different levels of microphysical description (pure
  hydrodynamics, \texttt{HY} or leakage with only cooling, \texttt{LK}, or with
  heating/absorption included, \texttt{M0}). The abundance pattern for elements
  with $A\gtrsim 120$ is very robust and in overall good agreement with the
  Solar r-process abundances taken from \citet{arlandini:99}.} \label{fig:ejecta.yields}
\end{figure*}

\begin{figure}
  \includegraphics[width=\columnwidth]{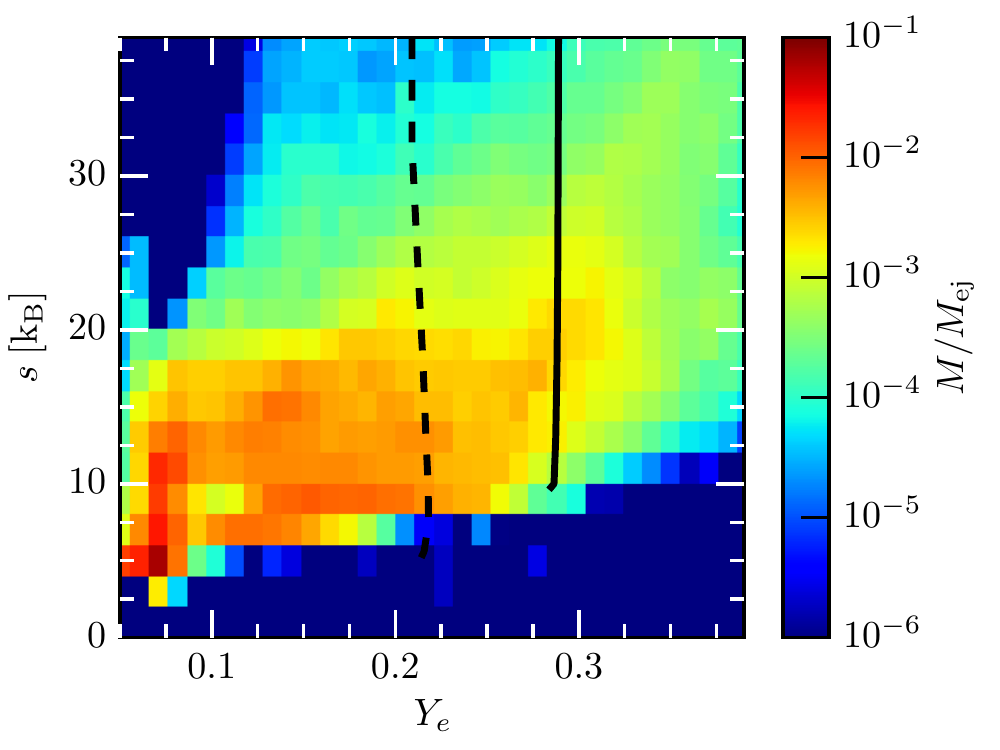}
  \caption{Joint distribution of composition and entropy for the \texttt{M0\_QC}
  simulation. The material to the left of the dashed contour line produces
  second and third r-process nucleosynthesis, while material to the
  right of the solid contour line only produces first peak r-process
  nucleosynthesis. The figure also hints at the existence of a correlation
  between $Y_e$ and $s$ as larger proton fractions are typically found in
  combination with high entropies.}
  \label{fig:ejecta.corr}
\end{figure}

We show the results of this procedure in Fig.~\ref{fig:ejecta.yields}, where we
plot the relative abundances of different elements in the final composition of
the ejecta for three of our models (\texttt{RP7.5}, \texttt{RP10}, and
\texttt{QC}) and with three different levels of microphysical description
(\texttt{HY}, \texttt{LK}, and \texttt{M0}).  The dynamical ejecta from all our
simulations is neutron rich with mass-averaged electron fractions $\langle Y_e
\rangle \lesssim 0.2$ (see Fig.~\ref{fig:histograms}).  We show the joint
distribution of $Y_e$ and specific entropy per baryon $s$ for simulation
\texttt{QC\_M0} in Fig.~\ref{fig:ejecta.corr}.  There is an approximate
correlation between $Y_e$ and $s$, due to the fact that shock heated material
undergoes more weak processing. However, the critical electron fraction for
producing third-peak r-process elements is relatively insensitive to $s$ (see
the contour lines in Fig.~\ref{fig:ejecta.corr}).  The bulk of the ejecta lies
in a region of parameter space where a robust r-process will occur.  In fact, we
find that fission cycling occurs in most of the material.  As a result, we find
the relative abundances for $A \gtrsim 120$ to be robust and close to Solar,
regardless of the merger type (eccentric vs.~quasi-circular) and of the neutrino
treatment. The only minor difference we find is that eccentric mergers show a
slight increase in the production of third-peak region nuclei, due to their more
neutron-rich ejecta.

The yields for $50 \lesssim A \lesssim 90$, referred to here as the first-peak
region, show greater variability than the yields at larger $A$.  Material in
this atomic mass range is produced in ejecta with $Y_e \gtrsim 0.22$, where an
incomplete r-process occurs.  Material quickly builds up in the first and second
peaks and then gets left there due to early neutron exhaustion.  The amount of
material with $Y_e \gtrsim 0.22$ is sensitive to both the binary parameters and
the treatment of neutrinos.  Without weak interactions, in the \texttt{HY}
models, almost all of the ejected material undergoes a complete r-process for
all of the binaries.  In the models including electron and positron captures but
not neutrino captures (\texttt{LK}), there is less first peak region material
for the eccentric \ac{BNS} mergers, which have slightly more neutron rich ejecta
than the quasi-circular mergers.  The further inclusion of neutrino captures in
the \texttt{M0} models has the largest impact on the yields of the \texttt{QC}
model. This is likely due to its low average ejecta velocity compared with the
eccentric mergers.  Nonetheless, neutrino captures only have a moderate impact
on the final yields for all simulated binaries.

Because of the small amount of mass in the high $Y_e$ tail of our ejecta,
first-peak nuclei are underproduced with respect to the Solar abundances (when
normalizing to the second r-process peak).  Therefore, we cannot account for all
of the r-process yields in the dynamical ejecta of binary neutron star mergers.
This is in contrast with the results of \cite{wanajo:14}, who found first-peak
region nuclei to be produced in approximately Solar proportion.  Their
simulations yield a much wider distribution of $Y_e$ than ours and a larger
fraction of their ejecta undergoes an incomplete r-process.  The reasons for
this discrepancy are unclear, but they could be due to differences in the
\ac{EOS} or in the treatment of neutrinos.

\subsection{Electromagnetic Counterparts}
\label{sec:results.EM}
The energy released by the radioactive decay of r-process nuclei powers
transients in the optical or near-infrared band that could potentially be
discovered through \ac{GW} or \ac{SGRB} detection follow-up observations and by
untargeted transient surveys.

We use a simple analytical model developed by \citet{grossman:14} to describe
the basic features of the macronova emission that would be produced by the ejecta
from our simulations. \citet{grossman:14} estimate the time at which the optical
or near-infrared signal peaks as the time when the radiation diffusion timescale
is equal to the dynamical timescale of the ejecta
\begin{equation}\label{eq:time_rad_peak}
\begin{split}
  t_{\mathrm{peak}} =\ & 4.9\ %
    \left(\frac{M_{\mathrm{ej}}}{10^{-2}\ M_\odot}\right)^{1/2} \times \\
    & \left(\frac{\kappa}{10\ \mathrm{cm}^2\ \mathrm{g}^{-1}}\right)^{1/2}
     \left(\frac{\langle v_\infty \rangle}{0.1\ c}\right)^{-1/2}
    \mathrm{days}\,,
\end{split}
\end{equation}
where $M_{\mathrm{ej}}$ is the ejected mass, $\kappa$ is an effective opacity of
the ejecta, and $\langle v_\infty \rangle$ is the mass-averaged asymptotic
velocity of the outflow.  They also estimate the peak bolometric luminosity
assuming a simple power-law decay for the energy release by the radioactive
decay of r-process elements $\dot{\epsilon} = \dot{\epsilon}_0
(t/t_0)^{-\alpha}$:
\begin{equation}\label{eq:lum_rad_peak}
\begin{split}
  L =\ & 2.5 \times 10^{40}
     \left(\frac{M_{\mathrm{ej}}}{10^{-2}\ M_\odot}\right)^{1-\alpha/2} \times \\
    &\left(\frac{\kappa}{10\ \mathrm{cm}^2\ \mathrm{g}^{-1}}\right)^{-\alpha/2}
     \left(\frac{\langle v_\infty \rangle}{0.1\ c}\right)^{\alpha/2}
    \mathrm{erg}\ \mathrm{s}^{-1}\,.
\end{split}
\end{equation}
Finally, the effective temperature can be computed assuming the area of
the emitting surface to be $4\pi\left(\langle v_{\infty} \rangle t_\mathrm{peak}
\right)^2$, which yields \citep{grossman:14}
\begin{equation}\label{eq:temp_rad_peak}
\begin{split}
  T =\ & 2200
    \left(\frac{M_{\mathrm{ej}}}{10^{-2}\ M_\odot}\right)^{-\alpha/8} \times \\
    & \left(\frac{\kappa}{10\ \mathrm{cm}^2\ \mathrm{g}^{-1}}\right)^{-(\alpha + 2)/8}
    \left(\frac{\langle v_\infty \rangle}{0.1\ c}\right)^{(\alpha-2)/2}
    K \,.
\end{split}
\end{equation}
The limitations of this simple model are discussed in \cite{grossman:14} and the
most serious one is that it tends to overestimate both peak time and
luminosity compared to more sophisticated numerical treatments.

We report the estimated peak times, luminosities and effective temperatures for
the macronova emission from our simulations in Tab.~\ref{tab:models}. The values
we quote are obtained using $\kappa=10\ \mathrm{cm}^2\ \mathrm{g}^{-1}$ as the
fiducial opacity of the r-process material. This value gives results that are
consistent with those of more sophisticated Monte Carlo calculations with large
databases of lines \citep{barnes:13}. For the energy production from the
radioactive decay of the r-process elements we follow \citet{grossman:14} and
choose $\alpha = 1.3$.

As can be inferred from Tab.~\ref{tab:models}, the macronova timescales and
luminosities show significant variation between our simulations. Simulations
such as \texttt{LK\_RP5}, which result in very little ejecta, have macronovae
that peak on very short timescales (less than a day) and that are relatively
blue compared to those of quasi-circular binaries.  The \texttt{RP7.5} and
\texttt{RP10} models, which eject significantly more material than the
\texttt{QC} model, have macronova light curves that peak on a timescale of one
to two weeks. They are also characterized by lower effective temperatures and
higher intrinsic luminosities with respect to quasi-circular binaries. We also
find the basic macronova properties to be rather sensitive to the level of
microphysical description. For instance, the peak time for the quasi-circular
binary goes from $2.8$ days to $1.4$ days or $1.6$ days when including
neutrino cooling or neutrino cooling and heating.

The ejecta are also expected to lead to radio emission as they transfer their
kinetic energy to the surrounding interstellar medium and produce non-thermal
synchrotron emission \citep{nakar:11}. The timescale over which the ejecta are
decelerated is given by \citep{nakar:11}
\begin{equation}\label{eq:time_dec}
\begin{split}
  t_{\mathrm{dec}} =\ & 30
    \left(\frac{E_{\mathrm{kin}}}{10^{49}\ \mathrm{erg}}\right)^{1/3} \times \\
    &\left(\frac{n_0}{\mathrm{cm}^{-3}}\right)^{-1/3}
    \left(\frac{\langle v_\infty \rangle}{c}\right)^{-5/3}
    \mathrm{days}\,,
\end{split}
\end{equation}
where $n_0$ is the density of the interstellar medium. Note that equation
\eqref{eq:time_dec} may be overestimating the deceleration time, because it does
not account for asphericities and non-uniform velocities in the outflows
\citep{piran:13, margalit:15}. However, our goal is not to provide detailed
estimates of the radio light curve.  These are difficult to construct in the
light of the large uncertainties in the ejecta properties. Rather, we want to
look for systematic trends. The simplified treatment we adopt should be
sufficient for this purpose.

Plasma instabilities develop at the location of the moving blast wave,
generating magnetic field and accelerating the electrons into a power law
spectrum $N(E)\propto E^{-p}$ with exponent $p$.  At an observing frequency
$\nu_{\mathrm{obs}}$ higher than both the self-absorption and synchrotron peak
frequencies the radio fluence (\ie, the flux density per unit frequency) for a
source at a distance $D$ can be expressed as \citep{nakar:11}
\begin{equation}\label{eq:obs_flux}
\begin{split}
  F_\nu =\  & 0.3
    \left(\frac{E_{\mathrm{kin}}}{10^{49}\ \mathrm{erg}}\right)
    \left(\frac{n_0}{\mathrm{cm}^{-3}}\right)^{\frac{p+1}{4}}
    \left(\frac{\epsilon_B}{0.1}\right)^{\frac{p+1}{4}} \times \\
    &\left(\frac{\epsilon_e}{0.1}\right)^{p-1}
    \left(\frac{\langle v_\infty \rangle}{c}\right)^{\frac{5p-7}{2}}
    \left(\frac{D}{10^{27} \mathrm{cm}}\right)^{-2} \times \\
    &\left(\frac{\nu_{\mathrm{obs}}}{1.4\ \mathrm{GHz}}\right)^{-\frac{p-1}{2}}
    \mathrm{mJy}\,,
\end{split}
\end{equation}
where $\epsilon_{\mathrm{B}}$ and $\epsilon_{\mathrm{e}}$ are the efficiencies
with which the energy of the blast wave is transfered to the magnetic field and
to the electrons, respectively.

The properties of the radio remnant produced by the ejecta from our simulations
are reported in Tab. \ref{tab:models}. To compute $t_{\mathrm{dec}}$ and
$F_\nu$, we adopt $n_0 = 0.1\ \mathrm{cm}^{-3}$ as fiducial value for the
number-density of the interstellar medium, which is a reasonable value for a
\ac{GC} \citep{rosswog:13}. We also take $p=2.3$, $\nu_{\mathrm{obs}} = 1.4\
\mathrm{GHz}$, and $\epsilon_{\mathrm{B}}=\epsilon_{\mathrm{e}}=0.1$ following
\citet{nakar:11}.  Finally, we place our binaries at a reference distance of $D
= 3.086\times 10^{26}\ \mathrm{cm}\simeq 100\ \mathrm{Mpc}$.

As with the macronova emission, we find significant variation in the properties
of the radio transients emergent from our simulations. This is not surprising
given that the deceleration timescale and the radio fluence depend on the mean
velocity and on the kinetic energy of the ejecta, which in turn change with
model and neutrino treatment. The timescale for the radio emission varies from
$\sim 1$ year to almost $30$ years between different simulations. The radio
fluence varies by more than two orders of magnitude between all our runs, going
from $0.015\ \mathrm{mJy}$ in simulation \texttt{M0\_QC} to $2.170\
\mathrm{mJy}$ in simulation \texttt{HY\_RP7.5}. Deceleration timescale and
fluence are also strongly dependent on the microphysical treatment. For
instance, $F_\nu$ changes by almost a factor of ten (from $2.170\ \mathrm{mJy}$
to $0.274\ \mathrm{mJy}$) for the \texttt{RP7.5} model when switching on
neutrino cooling.

\section{Summary and Discussion}
\label{sec:conclusions}

In this study, we presented a number of full-\ac{GR} numerical simulations of
\ac{BNS} mergers employing a microphysical \ac{EOS}. We considered both mergers
of binaries in quasi-circular orbits and of eccentric binaries produced through
dynamical capture in dense stellar environments. We systematically varied the
level of our treatment of weak reactions in our simulations to isolate the
impact of neutrino emission and absorption on the composition and morphology of
the outflows.

We identify three main components of the dynamical ejecta in our simulations.  A
first component, driven by tidal interactions between the two \acp{NS}, is cold
and very neutron rich (with $Y_e \lesssim 0.1$), and it is confined within $\sim
20^\circ$ of the orbital plane. A second component is driven by shocks formed
during the merger and is less neutron rich with $Y_e \gtrsim 0.15$, but more
isotropic, being spread over an angle of $\sim 60^\circ$ from the orbital plane.
A third, relatively proton-rich, with $Y_e \gtrsim 0.25$, but very tenuous,
component is observed at high-latitudes, especially in simulations that include
neutrino heating. The relative importance of each of these components varies
between the simulations and is the result of the interplay between the bulk
dynamics and the effects of neutrino transport. All of the ejecta are
sub-relativistic with $\langle v_\infty \rangle \lesssim c/3$ for all but the
most extreme, nearly head-on, configurations, which, however, result in very
little ejecta, because of prompt \ac{BH} formation.

In agreement with previous work \citep{east:12b, rosswog:13}, we find that
eccentric binaries can eject orders of magnitude more mass than binaries in
quasi-circular orbits and only slightly less than \ac{BHNS} binaries. Somewhat
surprisingly, and different from Newtonian results \citep{rosswog:13}, we find
that the ejecta from eccentric mergers are typically more neutron rich than
those of quasi-circular mergers. The reason for this is that the ejecta in
eccentric mergers is increasingly dominated by the tidal component. This trend
is the opposite of what is found in Newtonian simulations by \citet{rosswog:13},
which found eccentric mergers to yield more proton-rich outflows than
quasi-circular mergers. The reason for this difference is that, on the one hand,
Newtonian calculations tend to overestimate the relative ratio of tidally- to
shock-driven ejecta for quasi-circular binaries \citep{bauswein:13}. On the
other hand, Newtonian studies also overestimate the amount of shocked ejecta for
eccentric binaries.

Our results also indicate that neutrino cooling and heating have an important
impact on the composition, morphology, total mass of the outflows, and remnant
torus masses.  Simulations performed neglecting neutrino cooling and heating
consistently overestimate ejecta masses by a factor of $\sim 2$ or more.  They
also over-predict the amount of mass ejected at high-latitudes by a factor of a
few.  Given the impact that baryonic pollution in the polar regions of \ac{SGRB}
engines may have \citep{just:15b}, this suggests that \ac{SGRB} engine studies
should use initial data from mergers simulations that included both cooling and
heating from neutrinos.

The effect of neutrino heating on the total unbound mass is comparably
smaller. However, we find the differences between simulations including or
neglecting neutrino heating to be significant enough to suggest that neutrino
heating should also be accounted for.

The absorption of neutrinos in the ejecta can also contribute to the
protonization of the outflow. We find this effect to be appreciable in our
simulations, although not as significant as reported by \citet{sekiguchi:15} and
\citet{foucart:15b}. The absorption of neutrinos seems to be affecting the
mass-distribution of the ejecta at high-$Y_e$, and it is especially significant
at high latitudes. At the same time, we find the variation in the average
electron fraction of the ejecta to be small and, in some cases, insignificant.
The differences between our results and those of \citet{sekiguchi:15} and
\citet{foucart:15b} could be due to the different treatments of neutrino
radiation and hydrodynamics, or of the methodology employed in the analysis. We
note that we find significant discrepancies in the composition of the ejecta
with \citet{sekiguchi:15}, even when comparing simulations that did not include
heating. Instead, we find good agreement with \citet{palenzuela:15}, who,
however, used a different nuclear \ac{EOS}.

We also find neutrino physics to have a significant impact on the basic
characteristics of the \ac{EM} counterparts from \ac{BNS} mergers. Switching on
neutrino cooling and heating can result in differences in the macronova peak
times of several tens of percent and up to factors of a few in some cases. Even
more drastic is the impact of the neutrino treatment on the properties of the
radio remnants created by the mergers. For instance, the radio fluence at $1.4\
\mathrm{GHz}$ for one of our models changes by almost a factor of ten when
neutrino cooling is switched on. This is a consequence of the fact that the
properties of the radio emission depend crucially on the asymptotic kinetic
energy of the ejecta, which, is affected by the cooling.

We estimate the nucleosynthetic yields of our simulations using the tabulated
yields of \citet{lippuner:15}. We find that, despite their very diverse nature,
all considered \ac{BNS} mergers (eccentric or quasi-circular) robustly produce
r-process elements with atomic mass number $A\gtrsim 120$, with relative
abundances close to the Solar r-process abundance distribution. At the same
time, our yields show a deficit at lower $A$ compared to the Solar abundances.
This is different from what was reported by \citet{wanajo:14}.  If confirmed,
our results would suggest that either core-collapse supernovae or late-time
neutrino, magnetically, and/or viscously driven winds \citep{dessart:09,
fernandez:13, siegel:14, metzger:14, fernandez:14, rezzolla:14, ciolfi:14,
fernandez:15, just:15a, martin:15, kiuchi:15a} would still be needed to produce
the least massive of the r-process nuclei.

We find the neutrino-driven ejecta at high latitudes, within $\sim 30^\circ$ of
the polar axis, to be relatively proton rich ($Y_e \gtrsim 0.3$) and mostly free
of lanthanides. This material will become optically thin on short timescales
\citep[\ie~less than a day,][]{kasen:13, tanaka:13}. This may have consequences
for the observation of macronovae counterparts to \acp{SGRB}, which are commonly
believed to be associated with mergers seen face-on. In the case of a long-lived
\ac{HMNS}, the ejecta from the disk surrounding the \ac{HMNS} may be
sufficiently proton rich to be also Lanthanide free \citep{kasen:15}, so that
the macronova will be dominated by the disk ejecta at early times, when the
dynamical ejecta close to the equatorial plane are still optically thick. In
this scenario, it may be possible to constrain the survival time of the
\ac{HMNS}, as suggested by \citet{kasen:15}. A possible uncertainty, however, is
related to the interaction between the disk and the dynamical ejecta close to
the equator, which may result in the disk wind ejecta being covered by the more
optically thick dynamical ejecta. This will need to be addressed by future
studies.

Important limitations of the present study are that we restricted ourselves to a
single \ac{EOS} and to equal mass binaries. We also ignored the effects of
\ac{NS} spin and of magnetic fields. Mass ejection may change quantitatively and
possibly qualitatively with different \ac{EOS} and with unequal mass binaries
\citep{rezzolla:10, hotokezaka:13, bauswein:13, sekiguchi:15, dietrich:15,
palenzuela:15, foucart:15b}, or with the inclusion of spin \citep{kastaun:15,
east:15b}. While the pre-merger magnetic fields are probably too weak to impact
the dynamical ejecta in quasi-circular binaries, this might change for eccentric
binaries undergoing multiple encounters. This is so because magnetic fields
might be significantly amplified by magnetohydrodynamical instabilities
triggered when the two \acp{NS} come into contact at their periastron
\citep{obergaulinger:10, zrake:13, kiuchi:15b}. In this way magnetic fields
could become dynamically relevant even before merger.  Finally, given the
qualitative and quantitative impact that neutrino radiation has on the ejecta,
it will be necessary to validate or replace currently employed neutrino
treatments with full multi-group (spectral) \ac{GR}
neutrino-radiation-hydrodynamics simulations.  Addressing these issues will be
object of future work.

\section*{Acknowledgements}
We thank S.~Bernuzzi, S.~Richers, and S.~Rosswog for useful discussions, and the
anonymous referee for comments that have improved the paper.  This research was
partially supported by the Sherman Fairchild Foundation, by NSF under award
nos.\ CAREER PHY-1151197, PHY-1404569, and AST-1333520, and by ``NewCompStar'',
COST Action MP1304. FG is supported by the Helmholtz International Center for
FAIR within the framework of the LOEWE program launched by the State of Hesse.
Support for LFR during this work was provided by NASA through an Einstein
Postdoctoral Fellowship grant numbered PF3-140114 awarded by the Chandra X-ray
Center, which is operated by the Smithsonian Astrophysical Observatory for NASA
under contract NAS8-03060. The simulations were performed on the Caltech compute
cluster Zwicky (NSF MRI-R2 award no.\ PHY-0960291), on SuperMUC at the LRZ in
Garching, on the NSF XSEDE network under allocation TG-PHY100033, on LOEWE in
Frankfurt, and on NSF/NCSA BlueWaters under NSF PRAC award no.\ ACI-1440083.

\bibliographystyle{mnras.bst}
\bibliography{references}

\appendix
\section{Neutrino Transport Details}
\label{sec:M0}
\subsection{The Boltzmann Equation for Free-Streaming Neutrinos}
We treat neutrinos as massless particles with four-momentum $p^\alpha$,
interacting with a (fluid) medium having four-velocity $u^\alpha$. Following
\citet{thorne:81}, we decompose $p^\alpha$ as
\begin{equation}
  p^\alpha = (-p_\beta u^\beta)(u^\alpha + r^\alpha)\,,
\end{equation}
where $E_\nu = - p_\alpha u^\alpha$ is the neutrino energy as measured by an
observer comoving with the fluid and $r^\alpha$ is a unit spacelike four-vector
orthogonal to the fluid four-velocity, \ie, such that
\begin{align}
  r_\alpha r^\alpha = 1\,, &&
  u_\alpha r^\alpha = 0\,.
\end{align}
From a more physical point of view, $r^\alpha$ represents the spatial direction
of propagation of the neutrinos as seen by an observer comoving with the fluid.
Finally, we introduce the four-propagation vector of the neutrinos
\begin{equation}\label{eq:radial.null}
  k^\alpha = u^\alpha + r^\alpha\,.
\end{equation}
Note that $k_\alpha k^\alpha = 0$.

The worldlines of neutrinos can be parametrized with the affine parameter
\begin{equation}
  \ell = \int (-p_\alpha u^\alpha) \dd s\,,
\end{equation}
so that
\begin{equation}
  \left(\frac{\partial}{\partial\ell}\right)^\alpha = k^\alpha\,.
\end{equation}
Using $\ell$, the Boltzmann equation for neutrino radiation transport can be
written as \citep{thorne:81}
\begin{equation}\label{eq:boltzmann}
  \frac{D F}{D \ell} = \mathbb{C}[F]\,,
\end{equation}
where $F$ is the distribution function of either electron or anti-electron
neutrinos and $\mathbb{C}$ describes the interaction between neutrinos and the
background fluid term \emph{in the fluid frame}. Finally, $D/D\ell$ is the total
derivative in phase-space along $p^\alpha$:
\begin{equation}
  \frac{D F}{D \ell} = k^{\alpha} \left[ \frac{\partial F}{\partial x^\alpha} -
     ?[c]\Gamma^{\delta}_{\alpha \beta}? p^\beta \frac{\partial F}{\partial
     p^{\delta}} \right] \,,
\end{equation}
where $?[c]\Gamma^{\delta}_{\alpha \beta}?$ are the Christoffel symbols.

\subsection{Neutrino Number Density Evolution}
Following an approach similar to that of \citet{Liebendorfer:09}, we decompose
neutrinos into a trapped component, which we treat with the leakage
prescription, and a free-streaming component, which we evolve using a moment
scheme. Our neutrino transport scheme evolves two equations describing the
streaming of neutrinos along radial rays and their average energy evolution.

It is easily seen that the free-streaming neutrino distribution function also
obeys the Boltzmann equation \eqref{eq:boltzmann}, with an appropriate
collisional term.  In the simulations presented here, this term is approximated
neglecting scattering of the free-streaming component of the neutrino radiation
and using the effective emissivity from the leakage scheme to compute neutrino
sources.  Finally, the absorption opacities of electron neutrinos and
anti-neutrinos are computed in the local thermodynamical equilibrium
approximation, as in our leakage scheme.

We introduce the neutrino number current for a given neutrino flavor $X$
\citep{lindquist:66}
\begin{equation}\label{eq:neutrino.current}
  J_{_X}^\alpha = \int F p^\alpha \frac{\dd^3 p}{-p_0}\,.
\end{equation}
$J_{_X}^\alpha$ is such that $n_{_X} = -u_\alpha J_{_X}^\alpha$ is the neutrino
number density in the fluid rest frame. Neglecting scattering, the balance of
absorption and emission of the free-streaming neutrinos can be derived from the
first moment of the Boltzmann equation \citep{thorne:81, shibata:12},
\begin{equation}\label{eq:neutrino.continuity}
  \nabla_\alpha J^{\alpha}_{_X} = R_{_X}^{\mathrm{eff}} - \kappa_{_X} n_{_X}\,.
\end{equation}
In the previous equation, $R_{_X}^{\mathrm{eff}}$ is the effective neutrino
emission rate, while $\kappa_{_X}$ is the absorption opacity.

Note that equation \eqref{eq:neutrino.continuity} is exact but not closed. In
order to solve it, it is necessary to construct a closure. In our simulations we
close the equations assuming that neutrinos are streaming radially at the speed
of light.  This assumption is justified by the fact that we are only treating
the free-streaming neutrinos.

Mathematically, this is equivalent to assuming
\begin{equation}
  J_{_X}^\alpha = n_{_X} k^\alpha,
\end{equation}
where $k^\alpha$ is a fiducial null vector, which we construct from equation
\eqref{eq:radial.null} by taking $r^\alpha$ to be a radial unit-vector
orthogonal to the fluid four-velocity $u^\alpha$. This means that we are
assuming the free-streaming neutrinos to be streaming radially in a frame
instantaneously comoving with the fluid.

Under this assumption, we can derive a balance equation for $n_{_X}$
\begin{equation}\label{eq:neutrino.continuity.approx}
  \partial_t (\sqrt{-g} n_{_X} k^t) + \partial_r (\sqrt{-g} n_{_X} k^r) =
    \sqrt{-g}\big( R_{_X}^{\mathrm{eff}} - \kappa_{_X} n_{_X} )\,,
\end{equation}
where $g$ is the determinant of the 4-metric (in spherical coordinates).

During the post-merger evolution, we solve equation
\eqref{eq:neutrino.continuity.approx} on a series of independent radial rays
using a first order, fully-implicit finite volume method. To handle the coupling
with the hydrodynamics, we interpolate quantities interpolated from/to our
standard Cartesian \ac{AMR} grid every timestep. For the results presented here,
we used 2048 radial rays uniformly spaced in latitude and longitude and a radial
resolution $\Delta r \simeq 244\ \mathrm{m}$.

\subsection{Neutrino Average Energy Evolution}
Since we are interested in modeling both composition and temperature changes due
to neutrino absorption, a way to compute the average neutrino energy is also
needed. For this purpose, we evolve the average (free-streaming) neutrino
energies under the approximation of stationary spacetime, that is we assume that
$t^\alpha := (\partial_t)^\alpha$ is a Killing vector. Under this assumption the
quantity $(-p_\alpha t^\alpha)$ is conserved along neutrino worldlines in
absence of interactions with the fluid:
\begin{equation}
  \frac{\dd(-p_\alpha t^\alpha)}{\dd\ell} = 0\,.
\end{equation}
The quantity $\mathcal{E}_{_X} = - p_\alpha t^\alpha$ is the energy of neutrinos
of flavor $X$ as seen by the ``coordinate observer'' (a non-physical observer
with four-velocity $t^\alpha$). In particular
\begin{equation}
  \mathcal{E}_{_X} = - p_\alpha t^\alpha = - E_{_X} k_\alpha t^\alpha
                  =: E_{_X} \chi\,.
\end{equation}
Within this approximation, we can write an equation for the average neutrino energy
\begin{equation}\label{eq:avg.neutrino.energy.1}
  \frac{\dd \mathcal{E}_{_X}}{\dd\ell} =
  \frac{R_{_X}^{\mathrm{eff}}}{n_{_X}} \Big( \chi
  \frac{Q_{_X}^{\mathrm{eff}}}{R_{_X}^{\mathrm{eff}}} - \mathcal{E}_{_X} \Big),
\end{equation}
where $Q_{_X}^{\mathrm{eff}}$ is the effective neutrino energy source (which we
again take from the leakage scheme). Assuming radial propagation, equation
\eqref{eq:avg.neutrino.energy.1} can be rewritten as
\begin{equation}\label{eq:avg.neutrino.energy}
  n_{_X} k^t \partial_t \mathcal{E}_{_X} +
   n_{_X} k^r \partial_r \mathcal{E}_{_X} =
  \Big( \chi Q_{_X}^{\mathrm{eff}} - \mathcal{E}_{_X} R_{_X}^{\mathrm{eff}} \Big).
\end{equation}
This equation is solved on the same grid as equation
\eqref{eq:neutrino.continuity.approx} using a fully-implicit upwind 1st order
finite-differencing method.

\section{Coupling Between Hydrodynamics and Neutrinos}
\label{sec:coupling}
Equations \eqref{eq:tmunu} and \eqref{eq:composition} are solved using an
operator split approach. We use a standard 1st order explicit time update to
treat the momentum deposition by neutrinos.  However, to avoid issues with the
heating/cooling and composition sources becoming stiff, we use a semi-implicit
update formula for the evolution of the conserved energy and proton-number
densities.

In the context of the operator split method, we need to update energy and number
densities by solving equations of the form
\begin{equation}\label{eq:operator.split}
  \frac{\dd u}{\dd t} = f\,,
\end{equation}
over a single time step. In equation \eqref{eq:operator.split}, $f$ is possibly
large, while $u$ is a positive quantity (being a number or energy density),
which we wish to maintain positive during the evolution. Note that in our
\ac{EOS} implementation we include all binding energy contributions to the
energy density in the definition of $m_b$. In this way the specific internal
energy per baryon and the evolved energy density are always positive quantities.
Possibly problematic cases are regions with strong cooling (\ie, $f \ll 0$ and
$u < |f| \Delta t$) or small proton fractions.

To update $u$ while keeping it positive in these cases, we define the quantity
$\theta = f/u$ and update $u$ using the following semi-implicit scheme
\begin{equation}\label{eq:semi.implicit}
  \frac{u^{k+1} - u^k}{\Delta t} = \theta^k u^{k+1}\,.
\end{equation}
In the previous equation, the superscript $k$ denotes a function at time $t = k
\Delta t$. It is easy to see that equation \eqref{eq:semi.implicit} implies
\begin{equation}
  u^{k+1} = \frac{u^k}{1 - \theta^k \Delta t}\,,
\end{equation}
which manifestly ensures positivity of the solution (note that $\theta$ is
negative in the problematic cases).

\acrodef{AMR}{adaptive mesh refinement}
\acrodef{BH}{black hole}
\acrodef{CMA}{consistent multi-fluid advection}
\acrodef{CCSNe}{core-collapse supernovae}
\acrodef{BNS}{binary neutron stars}
\acrodef{BHNS}{black-hole neutron stars}
\acrodef{HMNS}{hypermassive neutron star}
\acrodef{EM}{electromagnetic}
\acrodef{EOS}{equation of state}
\acrodef{GC}{globular cluster}
\acrodef{GR}{general-relativistic}
\acrodef{GW}{gravitational-wave}
\acrodef{NS}{neutron star}
\acrodef{SGRB}{short gamma-ray burst}
\acrodef{TOV}{Tolman-Oppenheimer-Volkoff}

\end{document}